\documentclass[final]{aa}  

\makeatletter
\let\linenumbers\relax
\let\switchlinenumbers\relax
\makeatother

\usepackage{graphicx}
\usepackage{float}
\usepackage{txfonts}
\usepackage{lipsum}
\usepackage{caption}
\usepackage{subcaption}         % necessary for continued figures, example in section 3
\usepackage{lscape}             % to rotate a single page table, example in appendix.
\usepackage{placeins}           % useful with \FloatBarrier, to keep 
\usepackage{natbib}
\bibpunct{(}{)}{;}{a}{}{,} % Formattazione specifica richiesta da A&A
                                
\usepackage[colorlinks=true, allcolors=blue]{hyperref}
\usepackage{enumitem}
\usepackage{multirow}
\usepackage{xcolor}

\begin{document}

%%%%%%%%%%%%%%%%%%%%%%%%%%%%%%%%%%%%%%%%
% if you use custom commands in your title,
% ensure to check your title when submitting!
%%%%%%%%%%%%%%%%%%%%%%%%%%%%%%%%%%%%%%%%
   \title{\textsc{The Three Hundred} project: Low Gas Fraction Galaxy Cluster properties and their environment}
   \titlerunning{The300 project: Low Gas Fraction Galaxy Cluster properties and their environment}

   %\subtitle{Subtitle}

%%%%%%%%%%%%%%%%%%%%%%%%%%%%%%%%%%%%%%%%
% Please separate each author with the \and command
%
% Please do not include ORCIDs next to author names.
% Only ORCIDs authenticated by individual authors in EDPS
% editorial system will be taken into account.
% ORCIDs included here will be removed.
%%%%%%%%%%%%%%%%%%%%%%%%%%%%%%%%%%%%%%%%

   \author{Francesco Guidi\inst{\ref{sap}}\fnmsep\thanks{E-mail: guidi.1920254@studenti.uniroma1.it}
        \and Marco De Petris\inst{\ref{sap},\ref{inafrm}}\fnmsep\thanks{E-mail: marco.depetris@uniroma1.it}
        \and Sara Santoni\inst{\ref{sap},\ref{uam}}
        \and Raphael Wicker\inst{\ref{sap}}
        \and Antonio Ferragamo\inst{\ref{nap}}
        \and Hippolyte Froget \inst{\ref{Francia}}
        \and Stefano Andreon\inst{\ref{brera}}
        \and Weiguang Cui\inst{\ref{uam}, \ref{ciaff}, \ref{UK}}
        \and Elena Rasia\inst{\ref{trieste},\ref{trieste inaf},\ref{usa}}
        \and Gustavo Yepes \inst{\ref{uam}, \ref{ciaff}}
        % Andreon?
        }
    %\authorruning{Guidi et al.}

   \institute{Dipartimento di Fisica, Sapienza Università di Roma, Piazzale Aldo Moro, I-00185 Roma, Italy \label{sap}
   \and{INAF-Osservatorio Astronomico di Roma, Via Frascati 33, I-00078 Monteporzio Catone, Italy} \label{inafrm}
   \and Departamento de Física Teórica, Facultad de Ciencias, Universidad Autónoma de Madrid, Módulo 8, E-28049 Madrid, Spain \label{uam}
    \and Dipartimento di Fisica "Ettore Pancini", Università degli Studi di Napoli "Federico II", Via Cintia 21, I-80126 Napoli, Italy \label{nap}
    \and Département de Physique, Université Paris-Saclay, 625 Rue Louis de Broglie, F-91190 Orsay, France \label{Francia}
     \and INAF–Osservatorio Astronomico di Brera, via Brera 28, 20121, Milano, Italy
 \label{brera}
   \and IFPU - Institute for Fundamental Physics of the Universe, Via Beirut 2, I-34014 Trieste, Italy \label{trieste}
   \and Centro de Investigación Avanzada en Física Fundamental (CIAFF), Facultad de Ciencias, Universidad Autónoma de Madrid, 28049 Madrid, Spain \label{ciaff}
   \and Institute for Astronomy, Royal Observatory, Edinburgh EH9 3HJ, UK \label{UK}
   \and INAF - Osservatorio Astronomico di Trieste, Via Tiepolo 11, I-34131, Trieste, Italy\label{trieste inaf}
   \and Department of Physics, University of Michigan, Ann Arbor, MI 48109, USA \label{usa}
   }

   \date{Received ...; accepted ...}

% \abstract{}{}{}{}{}
% 5 {} token are mandatory
 
  \abstract
   {Galaxy cluster samples based on X-ray and Sunyaev-Zel'dovich (SZ) observations are affected by selection biases. These catalogs preferentially include systems with high gas content and surface brightness. 
   %This could lead to an incomplete sample, excluding objects with depleted gas content, hereafter referred to as low-gas-fraction clusters (LGFCs).
   Excluding objects with depleted gas content, hereafter referred to as low-gas-fraction clusters (LGFCs), could lead to an incomplete sampling.}
   {We aim to investigate the abundance and the properties of the LGFCs population using \textsc{The Three Hundred} hydrodynamical simulations, focusing on the \textsc{Gadget-X} code. In particular, we plan to study the outliers in the $f_{\rm g,500} - M_{500}$ relation, the influence of the environment, and their behavior in key scaling relations, with particular focus on the Compton-$Y$ observable.} 
   {We analyze a sample of $N_{\rm tot} = 9858$ simulated galaxy clusters from \textsc{The Three Hundred}, in the redshift band $z \in [0;0.817]$. 
%   The LGFCs are selected through a statistical analysis of the best-fit $f_{g,500}-M_{500}$ relation. 
    The LGFCs are selected through a statistical analysis as outliers of the $f_{\rm g,500}-M_{500}$ relation. 
   To analyze the effect of the environment, we compare the gas density and temperature radial profiles of the LGFCs and the No-LGFCs population, and we discuss the differences found. Finally, we study how the temperature, entropy, and spherical Compton parameter scaling relations are affected by the inclusion of LGFCs.}
   {We find that in our simulations LGFCs are preferentially found at the low-mass end and their abundance increases toward low redshift. Gas and temperature radial profiles of LGFCs are characterized by lower concentrations of gas in the core regions and higher temperatures, suggesting a more diffuse and heated ICM.  This behavior is also reflected in the entropy scaling relation, where LGFCs are extreme positive outliers. Contrary to observations, the $Y_{\rm sph,500}$ values of LGFCs show no significant deviation from the general population. Nevertheless, we cannot rule out that these differences are partly driven by the mass incompleteness at the low-mass end and the environmental bias of our simulated sample.}
   %: they follow the same relation and do not represent negative outliers.}
   {}


   \maketitle

%%%%%%%%%%%%%%%%%%%%%%%%%%%%%%%%%%%%%%%%%%%%%%%%%%%%%%%%%%%%%%
\section{Introduction}
\label{Introduzione}
Galaxy clusters (hereafter GCs) are the most massive gravitationally bound structures in the Universe and play a key role in modern cosmology. Their abundance, thermodynamic properties, and internal structure encode information on the underlying cosmological model.
In particular, the halo mass function links the number density of GCs to their mass and redshift \citep{Press1974}. This function depends on key cosmological parameters, such as $\Omega_{\rm m,0}$, which quantifies the present-day total matter density in the Universe, and $\sigma_{8}$, that is the density fluctuation in a sphere with a comoving radius $R = 8 \ h^{-1}$ Mpc at $z=0$. This parameter is sensitive to both the normalization and the shape of the primordial power spectrum.
Within this framework, reliable mass estimates and complete number counts allow for a precise calibration of the halo mass function, leading to robust constraints on $\Omega_{\rm m,0}$ and $\sigma_{8}$, and thus on the power spectrum.

It is widely recognized that selection effects in X-ray surveys represent a major source of systematic uncertainty in the construction of GC samples. For instance, \cite{Pacaud2007}, analyzing the \textit{XMM-Newton Large-Scale Structure Survey}, demonstrated that a precise calibration of the selection function is vital to determine the evolutions of scaling relations. Due to the steepness of the halo mass function, overluminous systems are preferentially detected. Similarly, \cite{Vikhlinin_2009} found that even when using robust X-ray total mass proxies calibrated via \textit{Chandra} observations and weak lensing, accounting for the bias and the scatter in the luminosity-mass relation is mandatory to correctly derive the halo mass function.

Recent observational studies \citep{Andreon2017, Andreon2024, Andreon2025} have shown that X-ray observational samples offer a biased view of the cluster population. This bias occurs because, at a given mass, brighter than average clusters are easier to select, whereas fainter ones are easily missed. Similarly, Sunyaev-Zeldovich (SZ) surveys do not fully alleviate this bias because they also tend to miss fainter than average clusters. Therefore, it is of primary importance to have a sample that includes GCs with known mass regardless of their Intra-Cluster Medium (ICM) content. In such samples, the probability of inclusion of the cluster is independent of its X-ray or SZ luminosity, ensuring that no system is selected or excluded based on its ICM properties.

One way to address this issue is to employ low-background X-ray telescope, such as \textit{Swift} X-ray telescope \citep{Burrows2005} to select a low-surface brightness population of GCs, such as the XUCS sample (X-ray Unbiased Cluster Survey) \citep{Andreon2016}. This sample comprises 34 GCs in the nearby universe ($0.050 < z < 0.135$) extracted from the C4 catalog of \cite{Miller2005}. These clusters lie in regions of low Galactic absorption and have masses derived using the caustic technique \citep{Diaferio1997, Serra2010}, with member velocity dispersion  $\sigma_v > 500$ km/s. Their mass spans the range $13.5 < \text{log} \ M_{500}/\mathrm{M_\odot} \leq 14.5$, where $M_{500}$ represents the mass of the sphere of radius $R_{500}$ such as $M_\Delta=4\pi R^3\Delta\rho_{\rm cr}$, with $\Delta = 500$ and $\rho_{\rm cr}$ is the critical density of the universe at the considered redshift. %which represents the mass contained within a radius sphere $R_{500}$ corresponding to an overdensity $\Delta=500$\footnote{It is defined as the ratio between the average density inside a sphere of radius $R_\Delta$ and a reference density, here taken as the critical density at redshift $z$: $\Delta = \frac{\rho(r\leq R_\Delta)}{\rho_{c}(z)}$.}, span the range $13.5 < \text{log} \ M_{500}/\mathrm{M_\odot} \leq 14.5$.

Another way to avoid biases due 
%to the selection methods based on ICM 
ICM-based selection methods
is to have a gravity-selected sample (hereafter, GS sample). Such a selection has the advantage of not requiring corrections for selection biases. An example is the subsample of 13 GCs extracted by \cite{Andreon2025} from the larger catalog of \cite{Oguri2021}. This sample was produced using very deep observations using the Hyper Suprime-Cam as part of a Subaru Strategic Program \citep{Aihara2022} and contains objects identified as weak lensing peaks in the photometric redshift range $0.12 < z_{phot} < 0.4$.

\cite{Puddu2021} find that GCs having low X-ray surface brightness within an aperture $R_{\Delta}$ also exhibit low gas fraction within the same overdensity aperture. 
This peculiar population of GCs characterized by low gas fraction, missed by X-ray and SZ catalogs but detected in optically and gravitationally selected samples, is defined as low gas fraction clusters (hereafter LGFCs). In this context, \cite{Andreon2025} investigate the scaling relation between the Compton-$Y$ parameter and total mass within $R_{200}$ in the GS sample. They found that the clusters in the GS sample are systematically below the $Y-M$ relation derived by \cite{Andreon2014} and \cite{Planck2014_XX}, with the presence of extreme negative outliers i.e., GCs having a $Y$ value which deviates more than $-2\sigma$ from the expected value for its mass \citep[as shown in Figs. 5 and 6 of][]{Andreon2025}. Therefore, LGFCs stand out as outliers in both the $L_{\rm X}-M$ and $Y-M$ planes. 

 This unseen population of clusters can have an impact on the calibration of scaling relations, since omitting gas poor clusters could cause a shift in both the normalization and the slope of these relations. Furthermore, this could lead to an underestimation of the intrinsic scatter \citep{Andreon2019}. Scaling relations, which link cluster observables to their total mass, are typically modeled as power-laws and can be derived from fundamental theoretical principles \citep{Kaiser1986}. They have been extensively calibrated through both observations \citep{Arnaud2005, Pratt2009, Lovisari2020} and numerical simulations \citep{Borgani2004, McCarthy2017, Cui2018}. It is, therefore, important to characterize this population of X-ray faint galaxy clusters.

In this context, cosmological hydrodynamical simulations play a pivotal role to understand these outliers. For instance, \cite{Ragagnin_2022}, using the \textsc{Magneticum} \footnote{\href{https://www.nottingham.ac.uk/~ppzfrp/The300/}{http://www.magneticum.org/}} simulation, confirm a strong covariance between the X-ray luminosity of the low-X-ray-surface-brightness objects and their gas fraction at fixed mass, in agreement with the observational findings of \cite{Puddu2021}. They also find that the LSB$_{\rm X}$ and gas poor systems are structurally older and more concentrated. By comparing full-physics runs against non-radiative simulations, they exclude environmental effects as the primary cause of gas depletion, concluding instead that active galactic nuclei (AGN) feedback mechanisms are responsible for efficiently ejecting gas during their formation history.
 
 We investigate the properties of the LGFCs population analyzing \textsc{The Three Hundred} hydrodynamical simulations \citep{Cui2018}. The characterization of the properties of LGFCs, and how they populate standard scaling relations or whether they constitute significant outliers is essential to understand what physical mechanism (AGN feedback mechanisms or other non-gravitational processes) is responsible for their gas depletion. Within the same simulation framework, previous studies explored the evolution of stellar and gas mass fractions \citep{Cui2018}, as well as the modeling of hot gas fraction-total mass relation \citep{Rasia2025}. 
%Furthermore, the impact of the different AGN feedback models provided by \textsc{Gadget-X} and \textsc{Gizmo-Simba} on the gas density, temperature, metallicity, pressure and entropy is explored in \cite{Cui2022, Li2023}.

The paper is organized as follows. In Sect. \ref{Simulations}, we briefly describe the characteristics of the simulations analyzed in this work and the corresponding sample of objects. In Sect. \ref{Method of analysis}, we analyze the gas fraction-mass relation. We then outline the statistical procedure adopted to identify LGFCs at different redshifts. Finally, we compare our results with some X-ray and SZ observational surveys. 
The impact of the large-scale environment in which GCs reside is examined in Sect. \ref{Environment}. We study the gas distribution and thermodynamic properties of GCs to assess whether LGFCs exhibit distinctive features. Furthermore, we check on any difference in the number of filaments connected to LGFCs and to No-LGFCs, quantified by the connectivity parameter.   
In Sect. \ref{Scaling Relations}, we investigate the thermodynamic properties of LGFCs in terms of mass-weighted temperature and electron entropy. We then focus on the behavior of LGFCs with respect to the $Y-M$ scaling relation, comparing our results with the ones found by \cite{Andreon2025}.
In Sect. \ref{Impact on total mass inference}, we examine whether the uncertainties affecting total mass inference in LGFCs differ from those typically observed in GCs samples.
Finally, in Sect. \ref{Conclusions}, we summarize our main findings and present our conclusions.

%%%%%%%%%%%%%%%%%%%%%%%%%%%%%%%%%%%%%%%%%%%%%%%%%%%%%%%%%%%%%%
\section{\textsc{The Three Hundred} project and Dataset}
\label{Simulations}
\textsc{The Three Hundred} project\footnote{\href{https://www.nottingham.ac.uk/~ppzfrp/The300/}{https://the300-project.org}} \citep{Cui2018} consists of a set of cluster-scale simulations centered on a mass-complete sample of the 324 most massive GCs (with $M_{\rm vir} \geq 8\times 10^{14} \ h^{-1} \ \mathrm{M_\odot}$), identified at $z=0$ in the dark matter only MultiDark Planck 2 simulation MDPL2 \citep{MDPL2_2016}. %Each zoom-in r
Each region extends to a comoving radius of $15 \ h^{-1}$ Mpc and 
%The edges of the simulation boxes are degraded in multiple layers with lower mass resolution particles which provide the same tidal fields as the original simulation. 
%These regions 
was re-simulated including baryonic physics, implemented through different hydrodynamical codes: \textsc{Gadget-Music} \citep{Springel2005}, \textsc{Gadget-X} \citep{Springel2005, Beck2016}, \textsc{Gizmo-Simba} \citep{Cui2022}. The first adopts the classical Smoothed Particle Hydrodynamics formulation. Conversely, \textsc{Gadget-X} uses an improved, modern hydrodynamics scheme.
%featuring artificial thermal diffusion, time-dependent artificial viscosity, and high-order Wendland C4 interpolating kernels. These improvements serve to better follow the gas particles and to reduce the gas clumpiness instability.
The latter uses the GIZMO cosmological hydrodynamics code in its Meshless Finite Mass solver mode. 

The advanced input physics enables detailed investigations on cluster properties such as their profiles, substructures, baryon content, dynamical state, and ICM thermalization. Initial conditions, for each zoom-in region, have been generated using the fully parallel \textsc{Ginnungagap}\footnote{\href{https://github.com/ginnungagapgroup/ginnungagap}{https://github.com/ginnungagapgroup/ginnungagap}} code \citep{ggp}. A total of 128 redshift snapshots have been stored for each simulation from redshift $z = 17$ to 0.
The cosmological parameters assumed by the simulation are those derived from the 2015 \textit{Planck} data release \citep{Planck2016}. They are reported in \cite{Cui2018}, along with other relevant simulation parameters, such as the mass of the high resolution  Dark Matter (hereafter, DM) $m_{DM} = 12.7\times 10^{8} h^{-1} \mathrm{M_\odot}$ and gas particles $m_{gas} = 2.36 \times 10^{8} h^{-1} \mathrm{M_\odot}$, and the softening length for these two components. 

In this work, we focus on the \textsc{Gadget-X} code, which is an updated version of the \textsc{Gadget-2} code. 
%This is a parallel TreeSPH code capable of following a collisionless fluid with the N-body method and models the gas component using smoothed particle hydrodynamics. 
It implements radiative gas cooling and star formation \citep{Springel2003}, chemical stellar evolution \citep{Tornatore2007, Biffi2017, Biffi2018}, stellar feedback and AGN feedback, following the implementations described in \cite{Rasia2015}.
%Black holes (BHs) are seeded with mass $M_{BH} = 5\times10^6 \ h^{-1}\mathrm{M_\odot}$ at the center of halos when their mass exceeds $2.5\times 10^{11} \ h^{-1}\mathrm{M_\odot}$. Their growth proceeds through mergers or gas accretion \citep[see][]{Bassini2020}. The accretion onto supermassive black holes (SMBHs) follows the $\alpha$-modified Bondi prescription, restricted by the Eddington limit, where $\alpha$ is a parameter fixed at 10 or 100 for hot gas ($T> 5 \times 10^5 $ K) and cold gas ($T<5 \times 10^5 $ K), respectively \citep{Gaspari2015}. 
The efficiency of AGN feedback depends on both the gas accretion rate and the supermassive black holes mass, and operates in two different modes: a quasar mode and a radio mode. The first is characterized by high accretion rates into the black hole; it is the most violent form that is capable of expelling the gas from the halo. The second is characterized by lower accretion rates and is less violent. 
%Its role is to keep the gaseous atmosphere of the halo hot at present \citep{HlavacekLarrondo2022}.

\subsection{The GCs sample}
\label{The sample}
 Each halo in the simulation is identified using the Amiga Halo Finder\footnote{\href{http://popia.ft.uam.es/AHF/}{http://popia.ft.uam.es/AHF/}} \citep[][]{KnollmanKnebe2009}, and is centered at the position of the local maximum of the total (baryonic + DM) density distribution. It is worth nothing, however, that our selection is strictly limited to groups around rich clusters. Thus, while the sample is mass complete only at $z=0$ for the most massive systems, it remains uncomplete for low-mass objects and environmental-biased.

Throughout this work, we focus on the $\Delta = 500$ aperture, which can be roughly associated with the area typically probed by X-ray and SZ observations. The resulting catalog spans a wide mass range of $10^{13} < M_{500} / [h^{-1} \mathrm{M_\odot}] < 10^{15} $, from the group scale to massive galaxy clusters. First, we restrict the resulting sample, considering only objects free of low resolution particles with $M_{500} > 10^{13.5} h^{-1}\mathrm{M_\odot}$ in the redshift range $z \in [0;0.817]$.  Systems below this mass are not resolved with a sufficiently large number of particles. Therefore, they can be affected by numerical artifacts, such as poor sampling of the gas component and stochastic feedback processes. This selection ensures that their thermodynamic properties are computed using only well-resolved particles, guaranteeing the reliability of the derived quantities. Moreover, for each zoom-in region, we select objects located within a sphere with radius $R = 10 \ h^{-1}$ Mpc from the center of the resimulated box. This choice minimizes boundary effects, which can pollute the quality of the results. These effects can bias the estimation of gas fraction and other thermodynamic properties.
 
 In Table \ref{Sample 2}, we report the total number of objects $N_{\rm tot}$ and their median mass $M_0$ of each of the considered redshift bin.

\begin{table}[ht!]
    \centering
    \caption{Summary of the analyzed sample. We report the redshift, total number of cluster and median mass $M_0$ of the cluster samples.}
    \begin{tabular}{c c c c}
    \hline\hline
    Snapshot & $z$ & $N_{tot}$ & $M_0 \ [ 10^{13}h^{-1} \mathrm{M_\odot}]$\\
    \hline
        101 &  0.817 & 976 & $6.26$\\
       
        104 & 0.700 & 1137 & $6.75$\\
       
        107 & 0.592 & 1215 & $6.79$\\
      
        110 & 0.490 & 1256 & $7.14$\\
        
        113 & 0.394 & 1310 & $7.13$\\
       
        116 & 0.304 & 1325 & $7.26$\\
      
        123 & 0.116 & 1329 & $7.59$\\
        
        128 & 0 & 1310 & $7.33$\\

        \hline
    \end{tabular}
\label{Sample 2}
\end{table}

\section{Gas fraction and selection of LGFCs}
\label{Method of analysis}
In this section, we first analyze the gas fraction-mass relation. We reproduce the analysis of the gas fraction-mass relation described in \cite{Rasia2025}, to verify the consistency with their findings, since our sample is different extending to lower masses. %Specifically, we compare three different functional forms: a linear model, a parabolic model, and a logarithmic model, 
%the latter proposed by \cite{Vikhlinin2009} and \cite{Stanek2010}. 
%with the second and the third similar to those respectively proposed by \cite{Vikhlinin2009} and \cite{Stanek2010}.
We then select the LGFCs population through a statistical analysis.

\subsection{Gas fraction - Mass relation}
\label{fg-M relation section}
To evaluate the gas fraction, we compute the total gas mass by summing only gravitationally bound and non-star forming gas particles with temperature $T > 0.3$ keV. This threshold selects the gas particles that contribute to the soft X-ray and SZ emission, which represent the observational proxies for inferring the gas mass. For the total mass $M_{500}$, we take all the components into account, that is, DM, gas, stars, and BHs.
 
The gas fraction is given by the following:
\begin{equation}
    f_{\rm g,500} = \frac{M_{\rm g,500}}{M_{500}}
    \label{Gas fraction formula}
\end{equation}
where $M_{g,500}$ and $M_{500}$ are the gas mass and the total mass within $R_{500}$.

Following the approach of \cite{Rasia2025}, we verify that a parabolic model yields a better fit to the data than either linear or logarithmic models. We therefore adopt the following functional form:
\begin{equation}
        f_{\rm g,500} = A + B \ \mathrm{log}(M_{500}/M_0) + C \ \mathrm{log}(M_{500}/M_0)^2
        \label{Parabola model}
\end{equation}
 where $M_0$ represents the median mass value in each redshift, as reported in Table \ref{Sample 2}.
 
\begin{figure}[ht!]
    \centering
    \includegraphics[width=9.0cm]{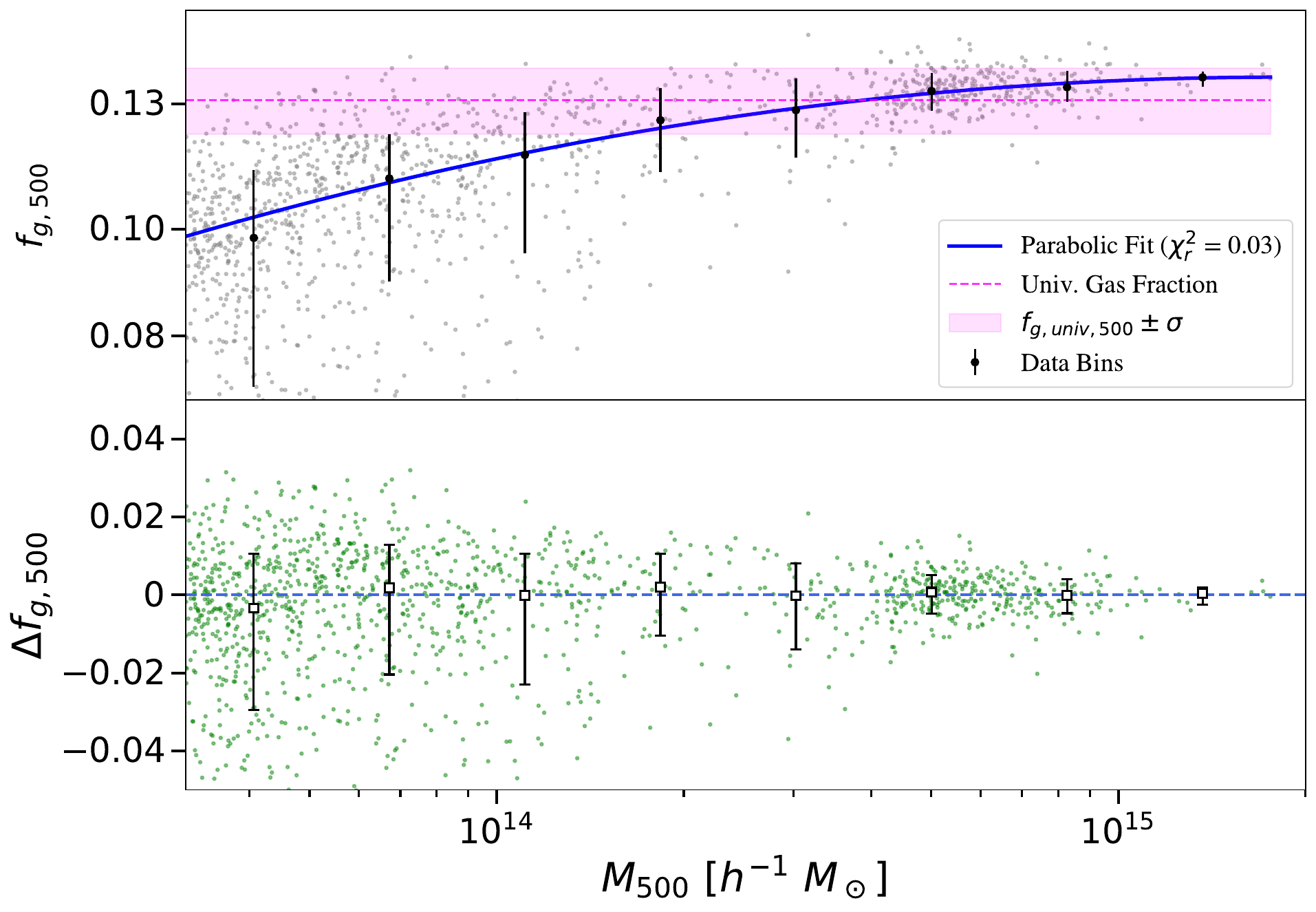}
    \caption{\textit{Upper panel}: $f_{\rm g,500} - M_{500}$ relation. The black error bars indicate the median values of $f_{\rm g,500}$ with the $16\mathrm{th}-84\mathrm{th}$ percentiles in every mass bin. The blue solid line is the best-fit to the data (Eq. \ref{Parabola model}). The magenta horizontal dashed line and its corresponding shaded area indicate the universal gas fraction value \citep{Eckert2019}. \textit{Lower panel}: Gas fraction residuals $\Delta{f}_{\rm g,500,i}$ for our GCs sample. The residual value for every GC in our sample at $z=0$ is plotted with a green dot. Black error bars represent the median and the $16\mathrm{th} - 84\mathrm{th}$ percentiles values in each mass bin, while the dashed blue line indicates the $\Delta{f}_{\rm g,500} = 0$ value.}
    \label{fgas e residui}
\end{figure}

As a representative example, the best-fit to the data at redshift $z=0$ is shown in upper panel of Fig. \ref{fgas e residui}. Each gray point represents one GC in our sample. The sample is divided into eight equally spaced logarithmic mass bins. The black error bars indicate the median values of $f_{\rm g,500}$ and the $16\mathrm{th}-84\mathrm{th}$ percentiles in every mass bin, and the solid blue line represents the best fit (Eq. \ref{Parabola model}). The horizontal dashed magenta line and its relative shaded area mark the universal gas fraction value $f_{\rm g,\mathrm{univ},500} = 0.131 \pm 0.009$ following the results found in \cite{Eckert2019}. This value is derived from a sample of local GCs ($0.04 < z <0.1$) and it is expected to have a negligible evolution down to $z=0$, thus we can use it as our reference value at $z=0$.
%We successively confirm the result by \cite{Rasia2025}, showing that the parabolic model provides the best fit for the data. This model is therefore adopted in all subsequent analyses.

%\begin{figure}[ht!]
    %\centering
    %\includegraphics[width=8.8cm]{fgas/fg_M500.pdf}
    %\caption{The black error bars indicate the %median values of $f_{\rm g,500}$ and the %$16\mathrm{th}-84\mathrm{th}$ percentiles in %every mass bin. The blue solid line is the %best-fit to the data (Eq. \ref{Parabola %model}). The magenta horizontal dashed line %and its corresponding shaded area indicate the %universal gas fraction value %\citep{Eckert2019}.}
    %\label{fgas profiles}
%\end{figure}

\subsection{LGFCs selection method}
\begin{figure}[ht!]
    \centering
    \includegraphics[width=8.8cm]{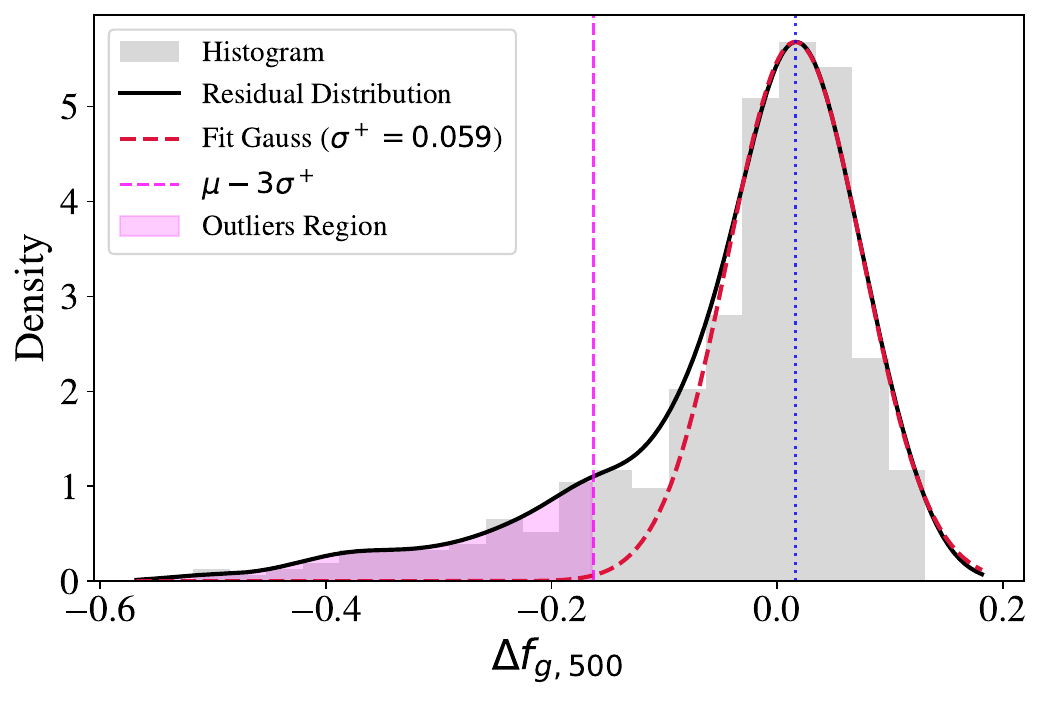}
    \caption{Residuals statistical analysis. The gray histogram represents the residual distribution, the black curve is a smoothed estimate of the distribution, and the red-dashed curve is the best-fitted Gaussian distribution $N(\mu,\sigma^+)$ obtained considering only the positive dispersion $\sigma^+$. The vertical magenta dashed line marks the $\mu -3\sigma^+$ threshold. The purple shaded area highlights the LGFC selection.}
    \label{Residual distribution}
\end{figure}

Using the best-fitting model identified in the previous section, for each object, we define the residual as the difference between its measured gas fraction value $f_{\rm g,500,i}$, and the corresponding value predicted by the model, $f_{\rm g,500,i}^{\rm fit}$:
\begin{equation}
    \Delta{f}_{\rm g,500,i} = f_{\rm g,500,i} - f_{\rm g,500,i}^{\rm fit}
\end{equation}
The selection of LGFCs is based on a statistical analysis of the residuals $\Delta{f}_{\rm g,500,i}$ of the gas fraction.
%As we show in Fig. \ref{Residual plot}, the residuals are then divided into $N=8$ mass bins, following the same binning scheme adopted in the previous plot. For each bin, we compute and display the median value and the $16\mathrm{th} - 18\mathrm{th}$ percentiles (black error bars). 
In the lower panel of Fig. \ref{fgas e residui}, we display them along with their median value and the $16\mathrm{th} - 84\mathrm{th}$ percentiles (black error bars) evaluated in mass bins. 
As a next step, we examine the distribution of residuals within each mass bin. In Fig. \ref{Residual distribution}, we show the results for the first (least massive) bin at redshift $z=0$ as a representative example. The residual distribution (black curve) exhibits a quasi-Gaussian shape only on the positive side and a long negative tail, indicating the existence of a population of GCs with lower gas content. 
%By definition, the median value of the residuals should be around zero, but not necessarily. 
%this is not the case. 
%The detected small deviations indicate local mismatches $\delta_j$ between the fit and the observed trend within each mass bin. Across the different mass bins, the values of the mismatch parameter $\delta_j$ range between  $-0.019$ and c. 
Small deviations of the median from zero indicate local mismatches $\delta_j$ between the fit and the observed trend, with values ranging from $-0.019$ to $+0.038$. 
Independently of the average value, we quantify the deviation from Gaussianity by fitting the positive side of each residual distribution with a Gaussian $N(\mu_j, \sigma^+_j)$ (red dashed curve) where $\mu_j$ is the peak residual value in the $j$-th mass bin (blue dotted vertical line) and $\sigma^+$ is the dispersion estimated using only positive residuals. Thus, we select LGFCs as those objects whose residuals satisfy the following:
\begin{equation}
    \Delta{f}_{\rm g,500,i} < \mu_j - 3\sigma^+_j.
\end{equation}
%i.e., systems lying more than $-3\sigma^+_j$ from $\mu_j$. 
This selection corresponds to the purple shaded area in Fig. \ref{Residual distribution}.

%\begin{figure}[ht!]
    %\centering
    %\includegraphics[width=8.8cm]{fgas/Res_z0.pdf}
    %\caption{Gas fraction residuals $\Delta{f}_{\rm g,500,i}$ for our GCs sample. The residual value for every GC in our sample at $z=0$ is plotted with an orange dot. Black error bars represent the median and the $16\mathrm{th} - 84\mathrm{th}$ percentiles values in each mass bin, while the dashed blue line indicates the $\mathcal{R}_{500} = 0$ value.}
    %\label{Residual plot}
%\end{figure}

\begin{figure}[h]
    \centering
    \includegraphics[width=8.8cm]{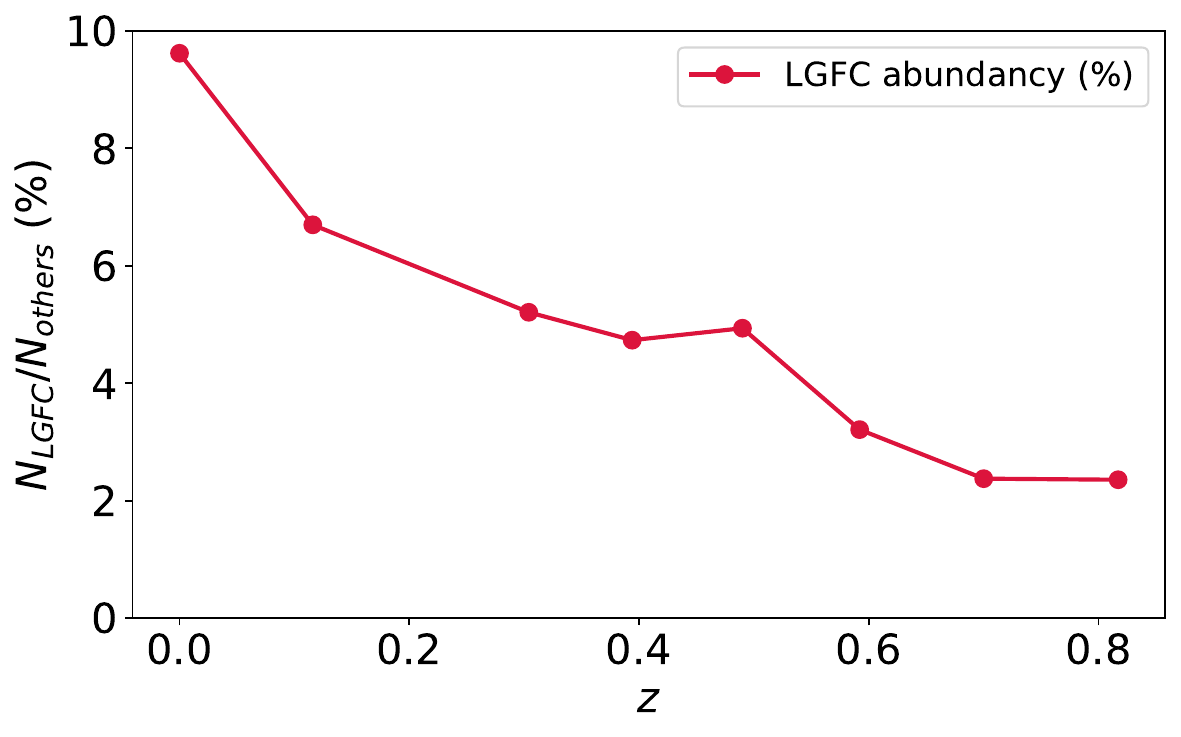}
    \caption{Redshift dependence of the LGFC population fraction.}
    \label{LGFC redshift evol}
\end{figure}

In Table \ref{LGFCs table}, we report the number of LGFCs, the total number of GCs, and the LGFCs fractional abundance $N_{\rm LGFCs}/N_{\rm tot} \ (\%)$ at each redshift. What is found is that LGFCs are preferentially found at the low mass end: only $ 1 \%$ of them have a $M_{500} \geq 10^{14.2} h^{-1} \mathrm{M_\odot}$. To disentangle whether this is a physical effect or a statistical artifact driven by the halo mass function, we analyze the fractional abundance of LGFCs across the entire sample. We find that this abundance decreases from $8\%$ at $M_{500} \sim 4 \times 10^{13} h^{-1} \mathrm{M_\odot}$ down to $0\%$ at $M_{500} \geq 4 \times 10^{14} h^{-1} \mathrm{M_\odot}$. This trend indicates that LGFCs are favored in lower mass systems. In Fig. \ref{LGFC redshift evol}, the evolution with redshift of their fractional abundance is shown. The LGFCs population grows toward lower redshifts, suggesting that gas depletion becomes progressively more pronounced over cosmic time.

\begin{table}[h]
    \centering
    \caption{Redshift evolution of the number of LGFCs, of the total number of GCs and of the relative abundance of LGFCs.}
    \begin{tabular}{c c c c}
    \hline\hline
    $z$ & $N_{\rm LGFCs}$ & $N_{\rm tot}$ & $N_{\rm LGFCs}/N_{\rm tot} (\%)$\\
    \hline
         0.817 &  23 & 976 & $2.36 \%$\\
         
         0.700 & 27 & 1137 & $2.37 \%$\\
         
         0.592 & 39 & 1215 & $3.21 \%$\\
         
         0.490 & 62 & 1256 & $4.94 \%$\\
         
         0.394 & 62 & 1310 & $4.73 \%$ \\
         
         0.304 & 69 & 1325 & $5.21 \%$\\
         
         0.116 & 89 & 1329 & $6.70 \%$\\
         
         0 & 126 & 1310 & $9.62 \%$\\
         \hline
         
    \end{tabular}
    
    \label{LGFCs table}
\end{table}

\subsection{Comparison with observational data}
\begin{figure}[ht!]
    \centering
    \includegraphics[width=8.8cm]{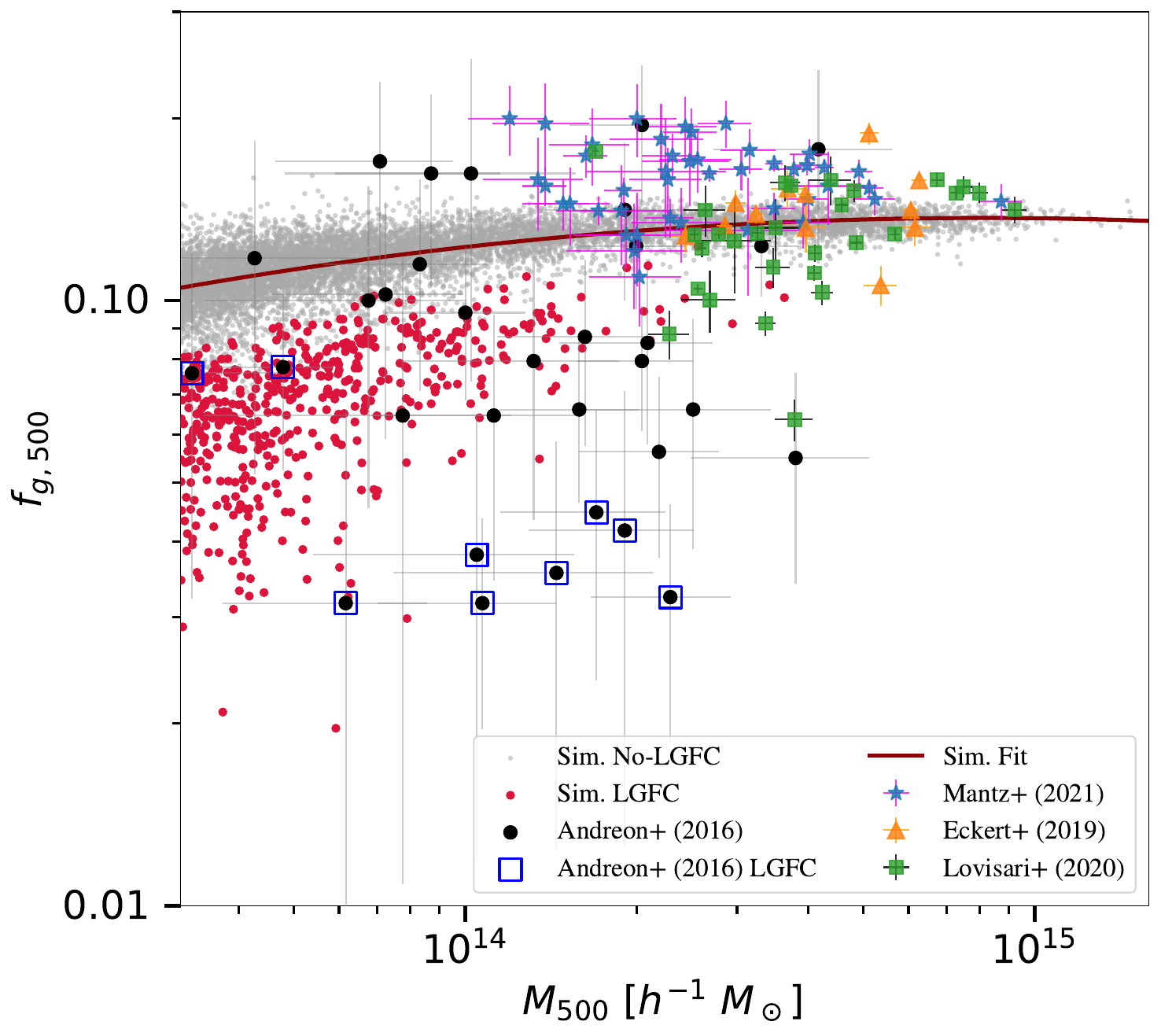}
    \caption{Comparison of our $f_{\rm g,500} - M_{500}$ derived from the simulation with observational datasets. Simulated data are reported as gray (No-LGFC population) and red (LGFC population) points. The brown curve is the best-fit relation (Eq. \ref{Parabola model}) on the full simulated sample. The black circles are the XUCS sample data \citep{Andreon2016}, with blue squares indicating their LGFCs. The data of \cite{Eckert2019}, \cite{Lovisari2020}, and \cite{Mantz2021} are plotted with orange triangles, green squares with black error bars, blue star points with magenta error bars, respectively.}
    \label{fgas Obs data comparison}
\end{figure}

To validate the goodness of our simulated data, we compare the properties of our GCs population with publicly available observational datasets. In Fig. \ref{fgas Obs data comparison}, we show the full simulated sample, combining all redshifts, which is composed of $N_{\rm tot}=9858$ objects. Gray and red points are, respectively, the No-LGFCs and the LGFCs population. The brown curve is the best-fit relation (Eq. \ref{Parabola model}). 
We choose three observational datasets to understand how different selection methods impact the observed scatter within the $f_{\rm g,500} - M_{500}$ plane. In this way, we can trace how specific strategies to select GCs either unveil or hide the LGFCs population.
The observational XUCS sample data of \cite{Andreon2017} are plotted with black circles. The points surrounded by blue squares are the clusters defined as LGFCs in that work.
The orange triangles correspond to the dataset of \cite{Eckert2019}. It contains 13 massive GCs ($3 \times 10^{14}\mathrm{M_\odot} <M_{500}<1.2\times 10^{15}\mathrm{M_\odot}$) in the nearby Universe ($0.04 < z < 0.1$), selected from \textit{XMM-Newton} and \textit{Planck} surveys in the Planck Sunyaev-Zel'dovich (PSZ1) source catalog \citep{PlanckSZ2015}. This sample is biased toward high X-ray surface brightness systems and its data points overlap well with our simulated data at the high-mass end. Only one GC with $M_{500} = 5.36 \times 10^{14}  \ h^{-1} \mathrm{M_\odot}$ has a gas fraction value that differs, at fixed mass, more than $-3.0 \sigma$ from the expected simulation value, where $\sigma$ represents the local intrinsic scatter of the simulation data at that specific mass scale, reaching a departure of $-5.6 \sigma$.
The green squares with black error bars represent the \cite{Lovisari2020} data. This catalog is based on \textit{XMM-Newton} observations of clusters in the Planck Early Sunyaev-Zel'dovich (ESZ) sample \citep{Pratt_2010} spanning the redshift range $0.059 < z < 0.546$. These data show a substantial overlap with our simulated sample, with an exception of a single extreme outlier of $-12.2\sigma$ at $M_{500} = 3.8 \times 10^{14} \ h^{-1} \mathrm{M_\odot}$.
Finally, blue star markers with magenta error bars represent the observational data of \cite{Mantz2021}, based on \textit{Chandra} X-ray telescope observations of 40 massive clusters in the redshift band $0.018 < z <  1.160$ selected for their relaxed X-ray morphology.

Among the compared datasets, the XUCS sample contains clusters with extremely low gas fraction. Some systems deviate by more $-3.0\sigma$ from the expected simulated value, with extreme outliers at $-16.3\sigma$ and $-14.6\sigma$ corresponding to $M_{500} = 2.3\times10^{14} \ h^{-1} \mathrm{M_\odot}$ and at $M_{500} = 3.8\times10^{14} \ h^{-1} \mathrm{M_\odot}$, respectively. In contrast, the other observational samples show reduced scatter and lack low surface brightness clusters. As \cite{Andreon2016} already stressed, this difference is most likely driven by selection effects: while X-ray luminosity and relaxation selections performed by \cite{Eckert2019} and by \cite{Mantz2021} do not lead to a large dispersion on the gas fraction values, SZ-selected catalogs like \cite{Lovisari2020} are more inclusive, showing a larger scatter. However, SZ-selected samples still fail to capture the dispersion revealed by the XUCS sample. In addition, standard SZ selections remain limited by mass and signal-to-noise thresholds that naturally favor more massive structures. Consequently, they lack the sensitivity to explore low and intermediate mass objects ($M_{500} \lesssim 10^{14} h^{-1} \mathrm{M_\odot}$) where XUCS and our simulated data show the most severe cases of gas depletion.

\section{Analysis of the cluster properties and environment}
\label{Environment}
The outer regions of GCs provide the interface between the ICM and the surrounding large-scale structure. Several physical processes operate in these outskirts ($r \gtrsim 3R_{200}$), including the breakdown of hydrostatic, thermal, and ionization equilibrium. Nevertheless, due to the low gas density in these regions, observational measurements are very challenging. Despite this, analyzing these regions is necessary to understand the physics of the in-falling matter and of the IGM in the filaments \citep{Walker2019}.

In this section, we investigate the cluster environment through the gas distribution in the outskirts, whereas in Sect. \ref{Connectivity} we characterize it in terms of its connectivity with the surrounding cosmic web framework.

In this section, we compute the radial profiles of gas density $\rho_{\rm g}(r)$ 
and mass-weighted temperature $T_{\rm mw}(r)$. We aim to compare the profiles of GCs selected as LGFCs with those of No-LGFCs, to identify if there are any systematic differences between the two populations. Our focus is restricted to these two specific quantities because they are the fundamental components defining the thermal gas pressure $P_e(r)$, which drives the SZ effect. Moreover, we analyze the filamentary structure around the clusters in order to check whether LGFCs have systematic differences in terms of filaments globally connected to them. 

The radial profiles are computed by accounting for all gas particles within the radial interval range $0.05 < r/R_{500} <5$ from each GC center, encompassing both cold and star-forming gas. We exclude the innermost core regions, where numerical artifacts, as gravitational softening length and limited particle sampling, could affect the results. Similarly, we exclude the more extended regions to ensure a high-sampled environment and the upper limit of $5R_{500}$ ($\sim 3.5R_{200}$) captures most of the relevant physics before transitioning to the warm-hot intergalactic medium of cosmic filaments.

%To compute the radial profiles, we first divide the environment in 40 logarithmically spaced radial bins. Then we compute the radial distance from the center for each $i$-th gas particle. 
%To minimize mass driven effects, we divide the sample into six tight mass bins. This specific resolution represents the optimal choice to ensure that each mass interval is narrow enough to isolate systems with nearly identical total mass, while retaining a sufficient number of structures per bin to guarantee the statistical robustness of the analyses. Once individual profiles are computed, we perform a statistical stacking to compute the median profile for these mass bins, to obtain representative radial trends. 
To compute the radial profiles, we consider 40 logarithmically spaced radial bins. We then perform a statistical stacking to compute the median profile in six mass bins. This specific resolution represents the optimal choice to ensure that each mass interval is narrow enough to isolate systems with nearly identical total mass, while retaining a sufficient number of structures per bin to guarantee the statistical robustness of the analyses. 
Moreover, we also compute the ratio between the median profiles of LGFCs and No-LGFCs. This provides a direct quantitative measure of the differences between the two populations.

\subsection{Gas density profile}
\label{Gas density profiles}
The average gas density in each radial bin is given by the gas mass divided by the volume of the bin.

\begin{figure*}[ht!] 
    \centering
    \includegraphics[width=180mm]{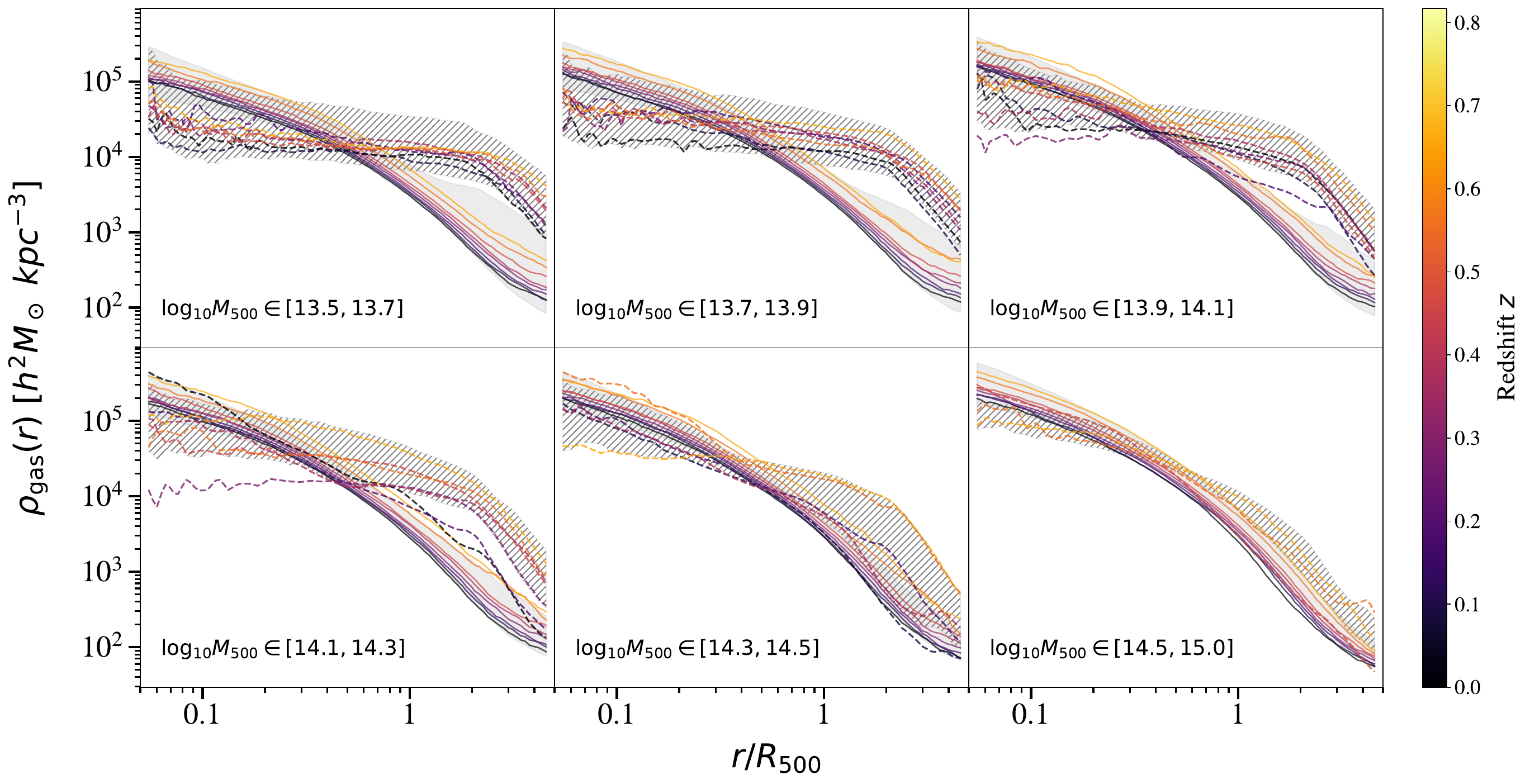}
    \caption{Gas density profiles $\rho_{\rm gas}(r)$ for six tight mass bins and for every redshift, plotted as a function of the normalized radius $r/R_{500}$. Solid lines are the profiles for the No-LGFCs, while the dashed profiles are those for the LGFCs population. Lines are colored by redshift value. Gray shaded region represents the intrinsic scatter for the No-LGFCs population, the hatching band those for the LGFCs population.}
    \label{Density profile}
\end{figure*}

In Fig.\ref{Density profile} we plot the median gas density profiles for the six mass bins. Solid lines are the median profiles for the No-LGFCs, while dashed lines are those for the LGFCs population. The shaded gray region around the solid curves shows the 16th - 84th percentile range for the No-LGFCs population, while the hatched region around the dashed curves represents the uncertainty range for LGFCs. Profiles are plotted as a function of normalized radius $r/R_{500}$ and are color-coded by redshift.

For the three lowest mass bins ($ 10^{13.5} \leq M_{500} / [h^{-1} \mathrm{M_\odot}] \leq 10^{14.4}$) LGFCs profiles are flatter than the others in the full radial range. This results in lower central densities ($r \lesssim 0.4 \ R_{500}$) and enhanced densities beyond $R_{500}$. Such behavior indicates a reduced degree of gas concentration possibly explained with an enhanced AGN feedback or inefficient gas accretion, which may redistribute gas from the core to larger radii. Alternatively, an environment bias due to the proximity of more massive systems could also account for this behavior.
In the most massive GCs, this difference disappears, with profiles becoming nearly indistinguishable between LGFCs and No-LGFCs. This is a sign that gravitational processes dominate over baryonic effects in high-mass systems.

Fig. \ref{Gas density ratio} shows the ratio between LGFCs and No-LGFCs median profiles (solid lines) with the respective uncertainty region (gray shaded area). The horizontal dashed line denotes the unit value. As discussed above, the main differences are in the three lowest mass bins. Near the core, the ratio is below unity, reflecting the behavior of the radial profiles just seen. At larger radii, a pronounced peak emerges around $r \sim 3R_{500}$, reaching values between 20 and 25.

\subsection{Gas temperature profile}
To study the thermal energy content of the ICM, we investigate the mass-weighted temperature profiles, as they provide a more direct measure of the gas thermal energy.
The mass-weighted gas temperature in each radial bin $j$ is defined as the sum of the products of each gas particle's mass and its electron temperature, normalized by the total gas mass contained within that same bin.

\begin{figure*}[ht!]
    \centering
    \includegraphics[width=180mm]{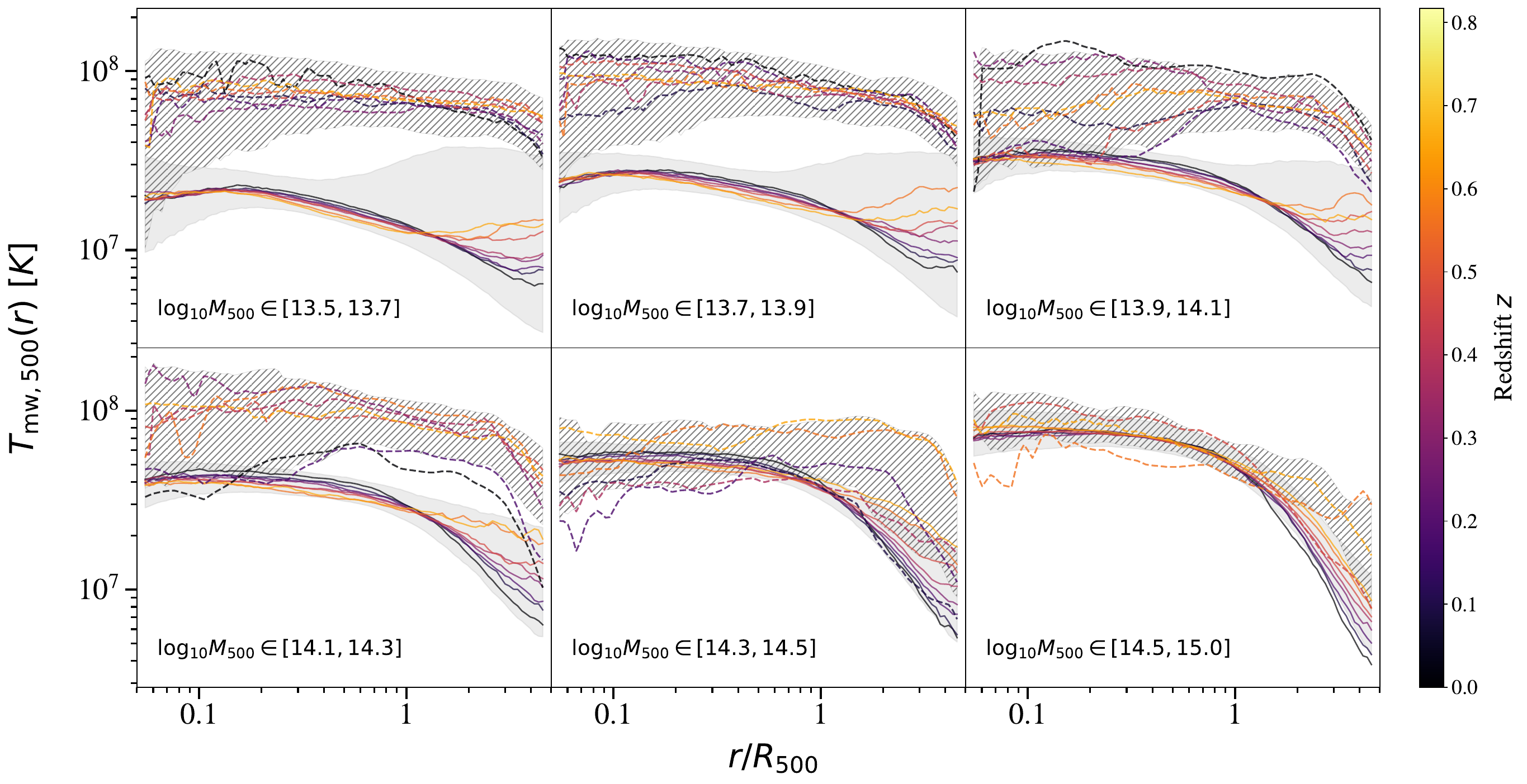}
    \caption{Same as Fig. \ref{Gas density profiles}, but now for gas temperature profiles $T_{\rm mw}(r)$. }
    %for six tight mass bins and for every redshift, plotted as a function of the normalized radius $r/R_{500}$. Solid lines are the profiles for the No-LGFCs, while the dashed profiles are those for the LGFCs population. Lines are colored by redshift value. Grey shaded region represents the intrinsic scatter for the No-LGFCs population, the hatching band those for the LGFCs population.}
    \label{Temperature profile}
\end{figure*}

In Fig. \ref{Temperature profile}, we plot the stacked profiles for the six mass bins. The solid lines are the median profiles for the No-LGFCs while the dashed lines are those for the LGFCs population, and they are colored by redshift value as for the density profiles. 

Fig. \ref{Gas temperature ratio} shows the ratio between the LGFCs and No-LGFCs median profiles. A strong mass dependence is clearly visible: in the mass bin $M_{500} \lesssim 10^{14.3} \ h^{-1} \mathrm{M_\odot}$, LGFCs exhibit higher temperature across the entire radial range, with the difference slightly growing toward the outskirts. The peak's amplitude decreases with increasing mass. Indeed, for the most massive systems (GCs with $M_{500} \gtrsim 10^{14.5} \ h^{-1} \mathrm{M_\odot}$) the temperature ratio remains close to unity at all radii, with minor deviations in the outskirts. This behavior is a confirmation of the interpretation we gave previously for the density, that is, the supremacy of gravitational processes over baryonic processes at higher masses due to their deeper potential wells. Moreover, it could be that our small mass systems, lying nearby massive systems, have a less steep profiles towards the outskirts due to the contamination of the environment.

By combining these results with density analysis it emerges that LGFCs are characterized by a lower gas density in the core region and a simultaneously higher temperature, implying a more diffused and heated gas. These effects decrease as the mass increases. This could be due to the stronger AGN activity (for the mass they have) in these systems.

\subsection{Environmental Connectivity of LGFCs}\label{Connectivity}
It is known that GCs reside at the nodes of the Cosmic Web and are connected by filamentary structures. In this framework, matter flows from lower-density toward higher-density regions, leading to the formation of compact GCs at the intersections between filaments \citep{Bond1996}. One way to quantify the filamentary skeleton around a GC is through the connectivity parameter, $k_{500}$, which is defined as the number of filaments connected to the GC within the sphere with radius $R_{500}$ \citep{Codis2018}. The properties of a GC are directly related by the number of filaments globally connected to it; in particular, the connectivity of a GC is strongly correlated with the total cluster mass \citep[see][Fig. 1]{Santoni2024}.  

\begin{figure*}[t!]
    \centering
    \includegraphics[width=160mm]{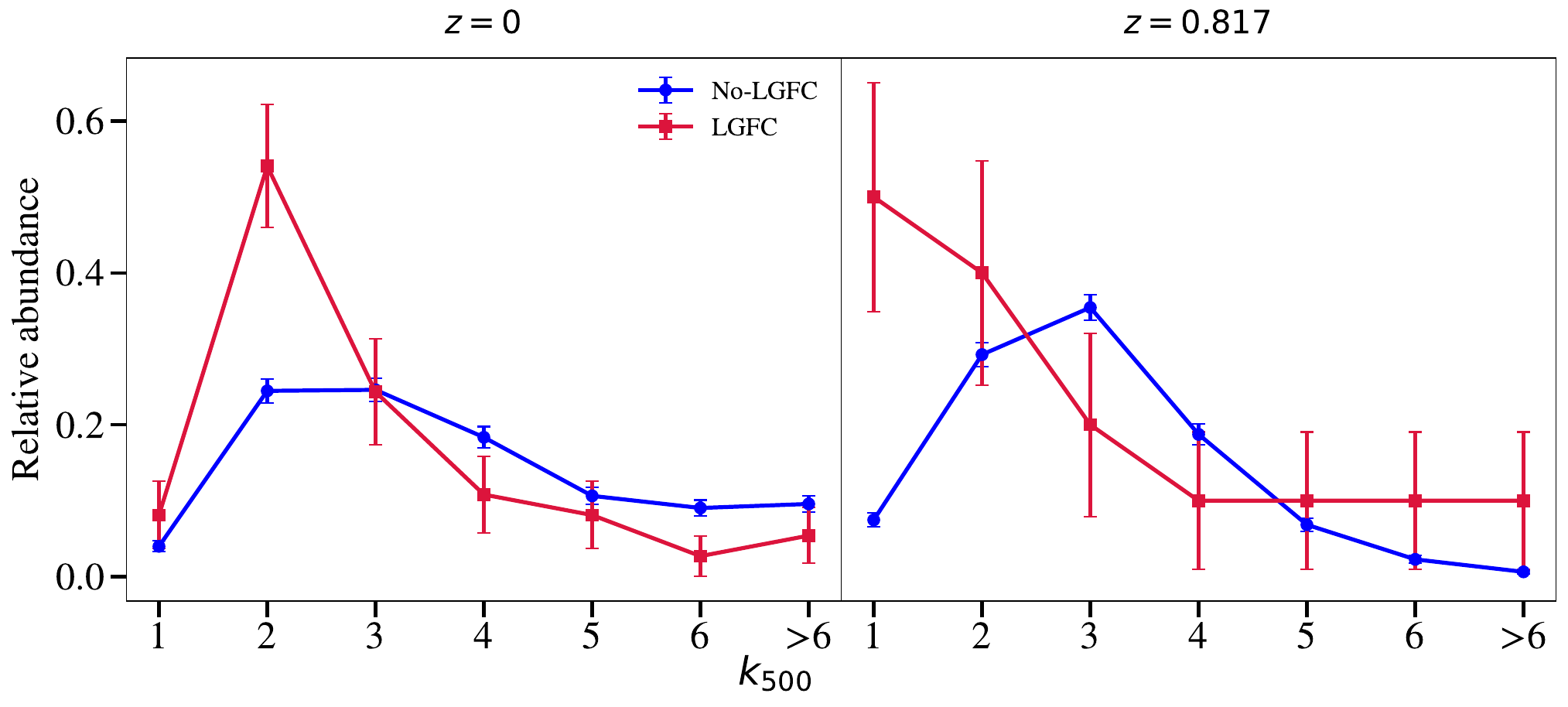}
    \caption{Connectivity $k_{500}$ distributions for LGFCs (red points) and No-LGFCs (blue points) at $z = 0$ (left panel) and at $z=0.817$ (right panel). Vertical error bars represent the $1 \sigma$ uncertainty derived by a Bayesian approach.}
    \label{Connectivity distributions}
\end{figure*}

In this section, we explore whether LGFCs differ in terms of connectivity with respect to the No-LGFCs, at the two extreme redshifts: $z = 0.817$ and $z=0$. Indeed, recent studies found indications that the position of GCs within the cosmic web could have an impact on their gas reservoir, with gas poor clusters rarely found closer to a filament \citep{Zarattini2025}. Specifically, a filament may in part control the gas flow: gas expelled from the central regions by AGN feedback may encounter resistance by the filament, reducing the speed and the distance that gas can reach.
The connectivity of clusters is estimated from the 3D gas filaments, identified with the cosmic web finder \textsc{DisPerSE} \citep{Sousbie2011}. We refer to \cite{Santoni2024} for the details of the filament identification and connectivity estimates of \textsc{The Three Hundred} clusters. 
%We use the connectivity estimated at $z=0$ for the GCs sample at $z=0$, while for the higher-redshift sample at $z=0.817$ we use the connectivity estimated at $z=1$, because it is the closest redshift for which the connectivity values have been estimated.
From this paper, we consider both the analysis of $z=0$ and $z=1$ which is the closest they have to the highest-redshift snapshot analyzed here ($z=0.817$).
In Fig. \ref{Connectivity distributions}, we show the distributions of $k_{500}$ values for the two GCs populations at $z=0$ (left panel) and at $z=0.817$ (right panel). We construct $7$ connectivity bins, combining all GCs with $k_{500} > 6$ into a single bin. This choice is made to ensure statistical robustness to our results, as there are a few GCs with $k_{500}>6$. Red points and blue points are the relative abundance for LGFCs and No-LGFCs, respectively, at each $k_{500}$ bin. The median values and the error bars indicate the posterior mean and the standard deviation derived from a Bayesian approach. For a bin containing $N_{\rm k}$ objects, the relative abundance and the standard deviation are given by: $p = (N_{\rm k} + 1)/(N_{\rm tot} + 2)$ and $\sigma_k =\sqrt{p(1-p)/(N_{\rm tot} + 3)}$, where $N_{\rm tot}$ is the number of the entire population.

At $z=0$, LGFCs tend to inhabit less connected zones of the Cosmic Web with only a single LGFC having $k_{500} > 6$. In contrast, the No-LGFCs population spans a wider range of connectivity. %, shifting toward higher connectivity values. 
The LGFC distribution peaks on $k_{500} = 2$ with a relative abundance value of $0.54$, while the No-LGFC distribution peaks at both $k_{500} =2$ and $k_{500} = 3$ with an abundance of $0.24$ in each bin.

At $z=0.817$ a similar qualitative behavior is observed. The LGFCs distribution peaks on $k_{500} = 1$ with a value of relative abundance of $0.5$, while the No-LGFCs population features a peak at $k_{500} = 3$ with an abundance value of $0.35$. However, the number of LGFCs in this redshift is limited, preventing any statistically robust conclusion.

These results suggest that the gas depletion observed in LGFCs may not be solely related to internal processes, but also to environmental ones. In less connected environments, the gas supply channeled into a cluster can be lower: when combined with internal feedback processes, this starvation effect can deplete the ICM, leading to a lower gas quantity.

\section{Do LGFCs follow $Y-M$ scaling relation?}
\label{Scaling Relations}
Scaling relations express how key physical quantities of GCs depend on their total mass. The theoretical framework is based on the self-similar model proposed by \cite{Kaiser1986}. Among other %the many 
approximations, this model assumes that clusters form from a purely gravitational collapse of the density peaks in the initial density field in an Einstein-de Sitter Universe. %Under these assumptions, c
Clusters are thus expected to be a scaled version of each other, following power-law relations derived from dimensional analysis, depending on the spherical overdensity considered.

Among all the scaling relations investigated to test this framework, one regards the Sunyaev-Zel'dovich effect \citep{Sunyaev1970}, which is the diffusion of cosmic microwave background photons within a hot plasma due to the inverse Compton scattering. It is one of the most powerful probes of cluster thermodynamics, both observationally \citep{PlanckSZ2015, Planck2016, Aguena_2026_ACTDR6}, and theoretically by the analysis of cosmological simulations  \citep{Sembolini_2013, LeBrun_2014, Dolag_2016}. Conversely to the X-ray surface brightness, which scales as $n_e^2$, the SZ signal depends on $n_e$, making it more suitable to study low density regions as cluster outskirts. Moreover, the SZ signal is redshift independent, in contrast to the X-ray surface brightness, which has a $(1+z)^4$ cosmological dimming.

The thermal SZ signal is characterized by the dimensionless Compton-$y$ parameter, which is defined as the integration of the thermal pressure of electrons present along the observer's line of sight:
\begin{equation}
    y = \frac{\sigma_T}{m_{\rm e}c^2}\int{P_{\rm e}(r)dl} = \frac{\sigma_{\rm T}k_{\rm B}}{m_{\rm e}c^2}\int{n_{\rm e}(r)T_{\rm e}(r)dl}
    \label{Compton y-parameter}
\end{equation}
where $\sigma_{\rm T}$ is the Thomson cross section, $k_B$ the Boltzmann constant, and $m_{\rm e}c^2$ is the electron rest energy.

Specifically, we focus on the spherically integrated quantity of the Compton parameter $Y$ within the $R_{500}$ aperture ($Y_{500,\rm sph}$), which is evaluated through the discrete form:
\begin{equation}
    Y_{500,\text{sph}} = \sum_{i \in R_{500}} y_i
    \label{Spherical Compton Parameter}
\end{equation}
where the index $i$ runs over all the contributions falling inside the 3D sphere of radius $R_{500}$.

The works carried out by \cite{Puddu2021, Andreon2024} and \cite{Andreon2025} suggest that LGFCs result in %turn out to have 
 lower values of the integrated Compton-$Y$ parameter than No-LGFCs with the same mass. This implies that the scaling relation between the $Y$ parameter and the total mass of GCs derived from ICM-selected samples may be biased, missing clusters with a low Compton $Y$ for their mass. This is an indication of the possible existence of a secondary population with a lower Compton-$Y$ parameter, which might comprises a significant fraction of objects, for example $13^{+10}_{-8} \ \%$ of the entire population in \cite{Andreon2025}. We therefore want to assess whether the same behavior is obtained in \textsc{The Three Hundred} simulated sample.
 
Before analyzing the $Y-M$ relation, we investigate how the mass-weighted temperature and the electron entropy evaluated inside $R_{500}$ depend on the total cluster mass $M_{500}$. This intermediate step allows us to disentangle the individual thermodynamic components which define the integrated Compton-Y parameter.

\subsection{$T_{\rm mw,500} - M_{500}$ dependence}
\label{T-M section}
\begin{figure*}[ht!]
\centering
    \includegraphics[width=160mm]{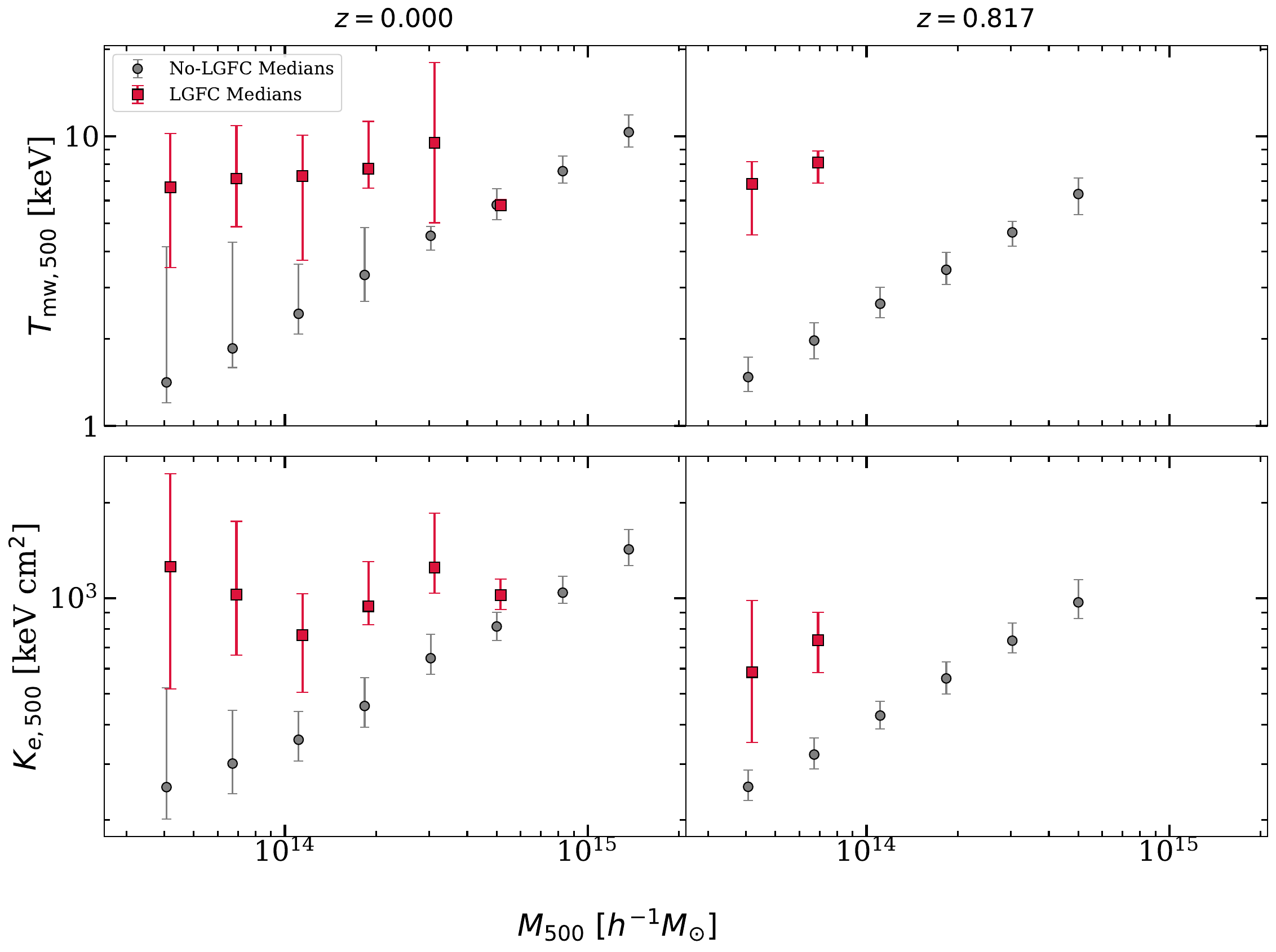}
    \caption{\textit{Upper panels}: $T_{\rm mw,500} - M_{500}$ trend. \textit{Bottom panels}: $K_{\rm e,500} - M_{500}$ trend. On the left-hand column we show the relations at redshift $z=0$, while on the right-hand column at redshift $z=0.817$. Gray points with errorbars refer to the No-LGFCs population medians and 16th-84th percentiles, while the red ones are those belonging to LGFCs only. For each mass bin, red points are slightly mass shifted for clarity.}
    \label{T-M scale law}
\end{figure*}

To study the thermal energy content of the ICM, we investigate the mass-weighted temperature profiles, as they provide a more direct measure of the gas thermal energy.
To compute $T_{\rm mw,500}$, we extend the temperature computation to the gas particles within the $R_{500}$ sphere. Then, we compare the temperature values between the No-LGFCs population and the LGFC population within the same logarithmically spaced mass bins used is Sec. \ref{fg-M relation section}. In the top panels of Fig. \ref{T-M scale law}, we present the $T_{\rm mw,500} - M_{500}$ dependence at the two redshift extremes. Within each mass bin, gray circles and error bars refer to medians and percentiles for the No-LGFCs population, while red points refer to those belonging to LGFCs only, slightly shifted along the mass axis for clarity.
At both redshifts and at fixed mass, LGFCs exhibit higher temperature. However, at $z=0$ the temperature enhancement of LGFCs is even more pronounced.
This redshift-dependence discrepancy supports the scenario in which non-gravitational processes (AGN feedback or environmental mechanisms) shape the thermodynamical properties of ICM. In low-mass systems, these processes can lead to fewer gas particles but the gravitational energy released during the LGFC formation is the same as other clusters with the same mass. Thus, the same energy is distributed among fewer particles, resulting in a higher energy per particle and higher temperatures.

% The consistently lower slope than the expected self-similar one, and the increase of such a discrepancy with decreasing redshift, support the scenario in which AGN feedback or environmental mechanisms play a crucial role in shaping the thermodynamical properties of ICM in GCs. In particular, non gravitational processes lead to fewer gas particles but the gravitational energy released during the LGFC formation is the same as other clusters with the same mass. Thus, the same energy is distributed among fewer particles, resulting in a higher energy per particle and higher temperatures.

Differently from the work of \cite{Ragagnin_2022}, using \textsc{Magneticum}, we could be affected by a potential environmental bias inherent to our simulation. Smaller mass clusters in our sample are not isolated systems, but they live in proximity to massive haloes.

\subsection{$K_{\rm e,500} - M_{500}$ dependence}
The entropy traces the thermodynamic history of the ICM, as it is primarily generated by shocks as gas accretes into the cluster potential well. It is, therefore, indicative of both gravitational (in the external regions) and non-gravitational (in the core) processes affecting the ICM \citep{Pratt2006}. We compute the electron entropy $K_{\rm e,500}$ by integrating within $R_{500}$ its radial profile, given by:
\begin{equation}
    K_{\rm{e}}(r) = n_{\rm{e}}(r)^{-2/3}k_{\rm B}T_{\rm mw}(r) = \bigg(\frac{\rho_{\rm g}(r)}{\mu_{\rm e}m_{\rm p}}\bigg)^{-2/3}k_{\rm B}T_{\rm mw}(r)
\end{equation}
where $n_e(r)$ is the radial profile of electron density, $\mu_{\rm e}$ is the mean molecular weight per free electron 
%(assuming a primordial composition of the ionized plasma with hydrogen fraction $X = 0.76$), 
and $m_{\rm p}$ is the proton mass.

In the bottom panels of Fig. \ref{T-M scale law}, we show the entropy-mass dependence at the two redshifts, $z=0$ and $z=0.817$. We adopt the same color coding, binning strategy, and symbols described in the previous section.
At both redshifts, the LGFCs population exhibits an entropy excess that becomes more pronounced in the $z=0$ case. This observed trend reflects those found in the temperature-mass dependence, given that the electron density is independent or very weakly dependent on the total mass.
 % The slope and normalization values for the electron density relation are reported in Table \ref{ne-M scale law table}, confirming a weakly dependence with the total cluster mass.

The elevated entropy observed in LGFCs arises from the combination of their reduced gas densities and high temperatures. Their entropy and temperature enhancement indicates that they represent a physically distinct population, whose properties cannot be explained within the self-similar framework alone.

 % \begin{table}[H]
 %     \centering
 %     \caption{$n_{\rm e,500} - M_{500}$ slope $B$ and normalization $A$ resulting from the two fits at redshift $z=0$ and $z=0.817$.}
 %     \begin{tabular}{c c c c}
 %     \hline\hline
 %          z & Population & A & B \\
 %          \hline
 %          0.817 & No-LGFCs & $-3.32 \pm 0.01$ & $0.02 \pm 0.01$\\
         
 %          0.817 & Total & $-3.31 \pm 0.01$& $0.01 \pm 0.01$\\
 %          \hline
 %         0 & No-LGFCs & $-3.26 \pm 0.01$& $0.06 \pm 0.01$ \\
        
 %          0 & Total & $-3.25 \pm 0.01$& $0.06 \pm 0.01$\\
 %     \hline
 %     \end{tabular}
 %     \label{ne-M scale law table}
 % \end{table}

\subsection{$Y_{500} - M_{500}$ scaling relation}
\begin{figure*}[ht!]
    \centering
    \includegraphics[width=160mm]{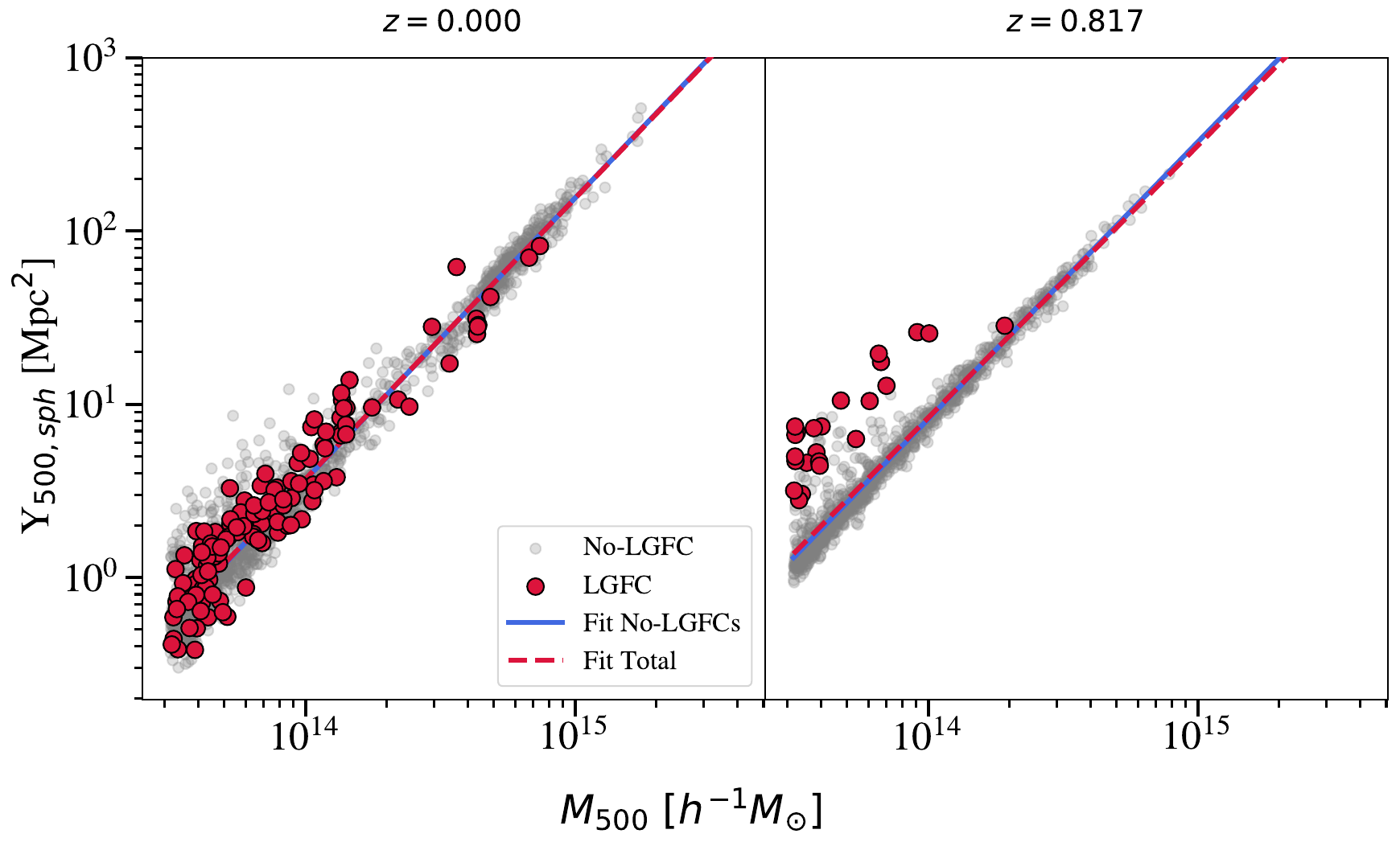}
    \caption{$Y_{\rm 500,\rm sph} - M_{500}$ scaling relation for our simulated data at $z=0$ (left panel) and $z=0.817$ (right panel). The blue solid line is the No-LGFCs only fit, the red dashed is the fit performed on the total population. Note that the two best-fit lines are overlapping.}
    %at the two extreme redshifts ($z = 0.817$ on the left side, $z = 0$ on the right). Gray points are No-LGFCs, while the red ones are LGFCs. The blue solid line is the best-fit considering only the No-LGFC population, while the red dashed line includes the entire population.}
    \label{Y-M scale law}
\end{figure*}

As discussed in Sect. \ref{Introduzione} and in Sect. \ref{Scaling Relations}, observational studies of the GS sample in \cite{Andreon2025} reveal the presence of GCs exhibiting remarkably lower values of the spherical Compton parameter at fixed mass, suggesting the presence of a population that is missed by ICM-selected surveys. Our goal is to observe whether a similar feature also emerges in our simulated sample.
In order to fit the $Y-M$ relation, we adopt the standard scaling relation model \citep{Kaiser1986}:
\begin{equation}
    Y_{500, \rm sph}(z) = 10^A E(z)^{2/3} \bigg(\frac{M_{500}}{M_0(z)}\bigg)^B
    \label{Scaling relation model}
\end{equation}
where $M_0(z)$ is the pivot value of the mass, assumed as the median mass of the sample at a given redshift (see Table \ref{Sample 2}) and $E(z)$ is the Hubble parameter evolution.
To quantify the impact of LGFCs in the $Y-M$ relation, we perform two different fits: one is performed considering only the No-LGFCs population, while the other accounts for the full sample. For clarity, we show plots and report best-fit parameters at the extremes of the redshift band analyzed, $z = 0.817$ and $z = 0$.
 In Fig. \ref{Y-M scale law}, we show the resulting plots at the two redshift extremes. In Table \ref{Y-M scale law table}, we report the best-fit parameters $A$ and $B$ of Eq. \ref{Scaling relation model}.

Although our LGFC subsample may be not fully representative of the observational samples, we do not find negative outliers with respect to the No-LGFCs population. This is in contrast with observational findings, where such a deficit is reported. Indeed, at $z = 0.817$, LGFCs exhibit a slight positive excess at fixed mass, that marginally flattens the fitted slope.
On the other hand, at $z = 0$, LGFCs follow the same relation as the No-LGFC population and do not represent positive or negative outliers, leaving the fit parameters unchanged.
This behavior is different from what is observed in the temperature-mass and entropy-mass relations. This discrepancy lies in the definition of the Compton parameter (see Eq. \ref{Compton y-parameter}). In the case of LGFCs, their lower gas density is compensated by their elevated gas temperature, yielding a thermal pressure comparable to that of No-LGFCs. As a consequence, the integrated SZ signal in LGFCs remains largely unchanged, compared to No-LGFCs. The result remains valid for the analyzed sample; however, its applicability to the observed cluster population may be limited by the mass incompleteness and environmental bias affecting the simulated LGFC subsample.

\begin{table}[H]
    \centering
    \caption{$Y_{500,\rm sph} - M_{500}$ slope $B$ and normalization $A$ resulting from the two fits at redshift $z = 0$ and $z=0.817$.}
    \begin{tabular}{c c c c}
    \hline\hline
         z & Population & A & B \\
         \hline
         0.817 & No-LGFCs & $0.59 \pm 0.01$ & $1.60 \pm 0.01$\\
         
         0.817 & Total & $0.60 \pm 0.01$& $1.58 \pm 0.01$\\
         \hline
         0 & No-LGFCs & $0.39 \pm 0.01$& $1.63 \pm 0.01$ \\
        
         0 & Total & $0.39 \pm 0.01$& $1.62 \pm 0.01$\\
    \hline
    \end{tabular}
    
    \label{Y-M scale law table}
\end{table}

\section{Impact on total mass inference}
\label{Impact on total mass inference}
The determination of the mass of GCs is of great importance, as it is directly related to the estimation of cosmological parameters through the halo mass function \citep{Press1974}, but it is not directly observable.
 The hydrostatic equilibrium (HE) mass is derived from the thermal pressure, gas density, and temperature profiles of the ICM. The HE mass can be computed through X-ray or SZ observations:
\begin{equation}
    M_{\rm HE,X}(<r) = -\frac{rk_{\rm B}T_{\rm g}(r)}{G\mu m_{\rm u}} \bigg[\frac{d \ \mathrm{ln} \ \rho_{\rm g}(r)}{d \ \mathrm{ln} \ r} +  \frac{d \ \mathrm{ln} \ T_{\rm g}(r)}{d \ \mathrm{ln} \ r}\bigg]
    \label{HE X mass}
\end{equation}
\begin{equation}
    M_{\rm HE,SZ}(<r) = -\frac{r^2}{G\rho_{\rm g}(r)} \frac{d  P_{\rm g}(r)}{dr}
    \label{HE SZ mass}
\end{equation}
where $P_{\rm g}(r)$ is the total thermal gas pressure profile, linked to $P_{\rm e}(r)$ through the relation $P_{\rm g}(r) = (\mu_{\rm e}/\mu)P_{\rm e}(r)$, where $\mu$ is the mean molecular weight, computed assuming a full ionized plasma, $G$ is the gravitational constant, and $m_u$ is the unit of atomic mass.

The hydrostatic mass bias $b$ quantifies the difference between the true mass of a cluster and the mass estimated under the assumption of HE:
\begin{equation}
    b = \frac{M_{\rm true} - M_{\rm HE}}{M_{\rm true}}
    \label{bias}
\end{equation}
where $M_{\rm true}$ is the true value of the mass of the GC. 

Given that the low gas fractions characterizing LGFCs may enhance the hydrostatic mass bias, we investigate whether the mass bias is correlated with the LGFCs population selected in this work.
To this end, we construct the radial profiles of gas density, temperature, and pressure, fitting them in the range $r/R_{500} \in [0.2, 3]$, as in \cite{Gianfagna_2022}. Also in our case, the innermost regions are excluded, because feedback and radiative cooling produce strong variations that are difficult to capture with smooth functional forms. 
The X-ray mass bias $b_{\rm X}$ (SZ mass bias $b_{\rm SZ}$) is then evaluated replacing $M_{\rm HE}$ with $M_{\rm HE,X}$ ($M_{\rm HE,SZ}$) in Eq. \ref{bias} at three overdensities: $\Delta = 2500, 500, 200$. In this case, we extend the redshifts range analyzed to $0.07 < z <  1.32$.
In Fig. \ref{bias redshift} we show the variation of $b_{\rm SZ}$ (left-hand panels) and of $b_{\rm X}$ (right-hand panels) along the redshift range and for the three overdensities here analyzed. Blue circles refer to the total population of GCs in our sample, red stars identify the bias values for LGFCs only, and green triangles indicate No-LGFCs only. The values of $b_{\rm X}$ and $b_{\rm SZ}$ for the No-LGFCs and LGFCs population are compatible along the redshift band: for all overdensities, the scatter within each redshift interval is comparable between the two populations.

In Fig. \ref{bias mass} we report the dependence of $b_{\rm SZ}$ and $b_{\rm X}$ on the total mass $M_{500}$. To increase the halo mass range, we simultaneously show the biases at four redshifts: $z=1.32$ (red points), $z=0.82$ (yellow points), $z=0.33$ (green points) and $z=0.07$ (blue points). At these four redshifts, LGFCs are marked with stars. The black dashed line and the shaded gray area represent the median and the 16th-84th percentile values of the biases at the mass bins. No dependence is visible: these biases are also independent on the total mass of the GC.

Overall, our results do not support the presence of a significantly different hydrostatic mass bias in LGFCs with respect to the full cluster sample. We can also come to this conclusion considering the temperature and density analyses made in the previous section. For LGFCs, the higher temperatures go in the direction of reducing the hydrostatic mass bias, but these are accompanied by almost flat gas density profiles, thus reducing the value of the gas density derivative. 

\section{Conclusions}
\label{Conclusions}
X-ray and SZ clusters samples could miss a population of GCs %having 
 with a significantly lower gas content than typical %structures 
 systems of similar mass, due to their reduced 
 ICM detectability. %surface brightness. Thus, these non-detections 
 Excluding these objects may probe a biased view of the cluster population. It is difficult to correct for this selection, as it relies on assumptions about the unseen population.

To investigate the peculiarities of the low gas fraction population, in this work we analyze \textsc{The Three Hundred} simulations focusing on the hydrodynamical re-simulations performed with the \textsc{Gadget-X} code.
The main findings of this work can be summarized as follows:
\begin{itemize}[label=\textbullet]

    \item[(i)] The population of LGFCs is predominantly composed of objects at the lower mass end  (only $1 \%$ of them have a $M_{500} \geq 10^{14.2} h^{-1} \mathrm{M_\odot}$) and their relative abundance to the total population grows as redshift decreases.

    \item[(ii)] The environmental properties of LGFCs are investigated by analyzing the radial gas density profiles, mass-weighted temperature profiles, and the distributions of the connectivity parameter $k_{500}$ and compared with the full population. At the low-mass end, LGFCs exhibit lower concentrations of gas in their central regions and simultaneously higher temperatures. In terms of connectivity, LGFCs tend to inhabit less connected environments.

    \item[(iii)] Examining the dependence regarding the gas mass-weighted temperature and the electron entropy with respect to the total GC mass we find that LGFCs exhibit, at fixed mass, higher values of gas temperature and entropy than the general cluster population. Analyzing the $Y_{500,\rm sph}-M_{500}$ relation, we find that the values of $Y_{500,\rm sph}$ for LGFCs do not differ with respect to the No-LGFCs having the same mass. 
    
    \item[(iv)] %The lack of full representativeness in our sample could affect
    Since our sample is not mass complete for low mass objects and environmental biased, the generalization of this analysis should be consider with caution as well as the comparison with the observational results in \cite{Andreon2016, Andreon2025}. In those studies, low gas fraction systems exhibit higher levels of gas depletion at a fixed mass than simulated systems, resulting in lower-than-expected SZ signals. Conversely, the work by \cite{Ragagnin_2022} does not suffer from such selection and environmental biases. This fundamental difference in the sample completeness and environmental distribution could potentially account for the observed discrepancies between our and their findings.

    \item[(v)] Finally, the analysis of the X-ray and SZ hydrostatic mass bias shows that this peculiar population exhibits consistent values of $b_X$ and $b_{SZ}$ across the entire redshift range, indicating that it does not introduce significant biases in the mass estimates. In addition, these biases are also total mass-independent.
\end{itemize}

These results suggest that LGFCs reside in less connected environments and they are characterized by a more diffuse and heated ICM: since the gravitational energy of LGFCs is the same compared to No-LGFCs with the same mass and is distributed among fewer particles, it results in an enhanced thermal energy per particle.
This is likely driven by stronger AGN feedback and other non-gravitational processes. These physical processes are more
efficient in low-mass systems, where the gravitational barrier
is shallower and less efficient to keep the gas tied up \citep{Puchwein2008, McCarthy2011, Barnes2017, Zhang2026}.

Interestingly, our findings are broadly consistent with those reported by \cite{Ragagnin_2022} from the Magneticum simulations, despite the different simulation framework and analysis strategy. Both studies converge on the interpretation that low-gas-fraction clusters are a physically distinct population whose gas depletion is primarily driven by non-gravitational feedback processes, likely associated with AGN activity. These systems are characterized by a reduced gas content and signatures of enhanced heating, including lower gas densities, higher temperatures and entropy compared to the general cluster population at fixed mass.

A possible future study will include the high-resolution version of \textsc{The Three Hundred}, simulated both with \textsc{Gadget-X} and \textsc{Gizmo-Simba} and considering also the BH/AGN properties, to control whether the impact of different feedback models affects the properties of LGFCs found in this work. Moreover, it would be valuable to track individual clusters across different epochs to investigate their evolution, providing a comprehensive overview of their gas evolution over time.

\begin{acknowledgements}
The simulations used in this paper have been performed in the MareNostrum Supercomputer at the Barcelona Supercomputing Center, thanks to CPU time granted by the Red Española de Supercomputación. As part of \textsc{The Three Hundred} project, this work has received financial support from the European Union’s Horizon 2020 Research and Innovation program under the Marie Skłodowska-Curie grant agreement number 734374, the LACEGAL project. FG, MDP, SS, AF, and RW acknowledge financial support from PRIN-MUR grant 20228B938N {\it"Mass and selection biases of galaxy clusters: a multi-probe approach"} funded by the European Union Next generation EU, Mission 4 Component 2 CUP B53D23004790006.
SS, WC and GY  would like to thank Ministerio de Ciencia e Innovación (Spain) for financial support under project grant PID2024-156100NB-C21. SS and WC  are supported by the Atracci\'{o}n de Talento Contract no. 2020-T1/TIC-19882 granted by the Comunidad de Madrid in Spain. SA acknowledges INAF grant  
{\it "Characterizing the newly  discovered clusters of low surface  brightness"} and PRIN-MUR grant
20228B938N {\it "Mass and selection biases of galaxy clusters: a multi-probe approach"}, the latter funded by the European
Union NextGenerationEU, Mission 4 Component 1  CUP C53D2300092 0006. 
\end{acknowledgements}

% --- BIBLIOGRAFIA ---
\bibliographystyle{aa}      % Utilizza il file aa.bst
\bibliography{bibliography} % Richiama il file bibliografia.bib (senza estensione)

@ARTICLE{ggp,
       author = {{Pilipenko}, Sergey and {Yepes}, Gustavo and {Gottl{\"o}ber}, Stefan and {Knollmann}, Steffen},
        title = "{Ginnungagap {\textemdash} A massively parallel cosmological initial conditions generator}",
      journal = {Astronomy and Computing},
         year = 2026,
        month = apr,
       volume = {55},
          eid = {101082},
        pages = {101082},
          doi = {10.1016/j.ascom.2026.101082},
archivePrefix = {arXiv},
       eprint = {2511.10353},
 primaryClass = {astro-ph.IM},
       adsurl = {https://ui.adsabs.harvard.edu/abs/2026A&C....5501082P}
}

@ARTICLE{Andreon2017,
  author  = {{Andreon}, S. and {Wang}, J. and {Trinchieri}, G. and {Moretti}, A. and {Serra}, A. L.},
  title   = {Variegate galaxy cluster gas content: Mean fraction, scatter, selection effects, and covariance with X-ray luminosity},
  journal = {\aap},
  year    = {2017},
  volume  = {606},
  pages   = {A24},
  doi     = {10.1051/0004-6361/201730722},
  month   = oct
}

@ARTICLE{Andreon2024,
       author = {{Andreon}, S. and {Trinchieri}, G. and {Moretti}, A.},
        title = "{Observed abundance of X-ray low surface brightness clusters in optical, X-ray, and SZ selected samples}",
      journal = {\aap},
         year = {2024},
        month = jun,
       volume = {686},
          eid = {A284},
        pages = {A284},
          doi = {10.1051/0004-6361/202345900},
       adsurl = {https://ui.adsabs.harvard.edu/abs/2024A&A...686A.284A}
}

@ARTICLE{Andreon2025,
       author = {{Andreon}, S. and {Radovich}, M.},
        title = "{Gravity-selected Cluster Samples: A New Take on the Compton-y--Mass Relation}",
      journal = {\apj},
         year = 2025,
        month = may,
       volume = {985},
       number = {1},
          eid = {78},
        pages = {78},
          doi = {10.3847/1538-4357/adc71e},
       adsurl = {https://ui.adsabs.harvard.edu/abs/2025ApJ...985...78A}
}

@ARTICLE{Andreon2016,
       author = {{Andreon}, S. and {Serra}, A.~L. and {Moretti}, A. and {Trinchieri}, G.},
        title = "{The amazing diversity in the hot gas content of an X-ray unbiased massive galaxy clusters sample}",
      journal = {\aap},
         year = 2016,
        month = jan,
       volume = {585},
          eid = {A147},
        pages = {A147},
          doi = {10.1051/0004-6361/201527408},
       adsurl = {https://ui.adsabs.harvard.edu/abs/2016A&A...585A.147A}
}

@ARTICLE{Burrows2005,
       author = {{Burrows}, D.~N. and {Hill}, J.~E. and {Nousek}, J.~A. and {Kennea}, J.~A. and {Wells}, A. and {Osborne}, J.~P. and {Abbey}, A.~F. and {Beardmore}, A. and {Mukerjee}, K. and {Short}, A.~D.~T. and {Chincarini}, G. and {Campana}, S. and {Citterio}, O. and {Moretti}, A. and {Pagani}, C. and {Tagliaferri}, G. and {Giommi}, P. and {Capalbi}, M. and {Tamburelli}, F. and {Angelini}, L. and {Cusumano}, G. and {Braeuninger}, H.~W. and {Burkert}, W. and {Hartner}, G.~D.},
        title = "{The Swift X-Ray Telescope}",
      journal = {\ssr},
         year = 2005,
        month = oct,
       volume = {120},
       number = {3-4},
        pages = {165-195},
          doi = {10.1007/s11214-005-5097-2},
       adsurl = {https://ui.adsabs.harvard.edu/abs/2005SSRv..120..165B}
}

@ARTICLE{Miller2005,
       author = {{Miller}, C.~J. and {Nichol}, R.~C. and {Reichart}, D. and {Wechsler}, R.~H. and {Evrard}, A.~E. and {Annis}, J. and {McKay}, T.~A. and {Bahcall}, N.~A. and {Bernardi}, M. and {Boehringer}, H. and {Connolly}, A.~J. and {Goto}, T. and {Kniazev}, A. and {Lamb}, D. and {Postman}, M. and {Schneider}, D.~P. and {Sheth}, R.~K. and {Voges}, W.},
        title = "{The C4 Clustering Algorithm: Clusters of Galaxies in the Sloan Digital Sky Survey}",
      journal = {\aj},
         year = 2005,
        month = sep,
       volume = {130},
       number = {3},
        pages = {968-1001},
          doi = {10.1086/431357},
       adsurl = {https://ui.adsabs.harvard.edu/abs/2005AJ....130..968M}
}

@ARTICLE{Diaferio1997,
       author = {{Diaferio}, A. and {Geller}, M.~J.},
        title = "{Infall Regions of Galaxy Clusters}",
      journal = {\apj},
         year = 1997,
        month = jun,
       volume = {481},
       number = {2},
        pages = {633-643},
          doi = {10.1086/304075},
       adsurl = {https://ui.adsabs.harvard.edu/abs/1997ApJ...481..633D}
}

@ARTICLE{Serra2010,
       author = {{Serra}, A.~L. and {Diaferio}, A. and {Murante}, G. and {Borgani}, S.},
        title = "{Measuring the escape velocity and mass profiles of galaxy clusters beyond their virial radius}",
      journal = {\mnras},
         year = 2011,
        month = mar,
       volume = {412},
       number = {2},
        pages = {800-816},
          doi = {10.1111/j.1365-2966.2010.17946.x},
       adsurl = {https://ui.adsabs.harvard.edu/abs/2011MNRAS.412..800S}
}

@ARTICLE{Puddu2021,
       author = {{Puddu}, E. and {Andreon}, S.},
        title = "{Do gas-poor galaxy clusters have different galaxy populations? The positive covariance of hot and cold baryons}",
      journal = {\mnras},
         year = 2022,
        month = mar,
       volume = {511},
       number = {2},
        pages = {2968-2976},
          doi = {10.1093/mnras/stab3632},
       adsurl = {https://ui.adsabs.harvard.edu/abs/2022MNRAS.511.2968P}
}

@ARTICLE{Andreon2019,
       author = {{Andreon}, S. and {Moretti}, A. and {Trinchieri}, G. and {Ishwara-Chandra}, C.~H.},
        title = "{Why are some galaxy clusters underluminous?: The very low concentration of the CL2015 mass profile}",
      journal = {\aap},
         year = 2019,
        month = sep,
       volume = {630},
          eid = {A78},
        pages = {A78},
          doi = {10.1051/0004-6361/201935702},
       adsurl = {https://ui.adsabs.harvard.edu/abs/2019A&A...630A..78A}
}

@ARTICLE{Planck2016,
       author = {{Planck Collaboration} and {Adam}, R. and {Aghanim}, N. and {Ashdown}, M. and {Aumont}, J. and {Baccigalupi}, C. and {Ballardini}, M. and {Banday}, A.~J. and {Barreiro}, R.~B. and {Bartolo}, N. and others},
        title = "{Planck intermediate results. XLVII. Planck constraints on reionization history}",
      journal = {\aap},
         year = 2016,
        month = dec,
       volume = {596},
          eid = {A108},
        pages = {A108},
          doi = {10.1051/0004-6361/201628890},
       adsurl = {https://ui.adsabs.harvard.edu/abs/2016A&A...596A.108P}
}

@ARTICLE{Cui2018,
       author = {{Cui}, W. and {Knebe}, A. and {Yepes}, G. and {Pearce}, F. and {Power}, C. and {Dav{\'e}}, R. and {Arth}, A. and {Borgani}, S. and {Dolag}, K. and {Elahi}, P. and others},
        title = "{The Three Hundred project: a large catalogue of theoretically modelled galaxy clusters for cosmological and astrophysical applications}",
      journal = {\mnras},
         year = 2018,
        month = nov,
       volume = {480},
       number = {3},
        pages = {2898-2915},
          doi = {10.1093/mnras/sty2111},
       adsurl = {https://ui.adsabs.harvard.edu/abs/2018MNRAS.480.2898C}
}

@ARTICLE{Springel2005,
       author = {{Springel}, V.},
        title = "{The cosmological simulation code GADGET-2}",
      journal = {\mnras},
         year = 2005,
        month = dec,
       volume = {364},
       number = {4},
        pages = {1105-1134},
          doi = {10.1111/j.1365-2966.2005.09655.x},
       adsurl = {https://ui.adsabs.harvard.edu/abs/2005MNRAS.364.1105S}
}

@ARTICLE{Rasia2025,
       author = {{Rasia}, Elena and {Tripodi}, Roberta and {Borgani}, Stefano and {Biffi}, Veronica and {Avestruz}, Camille and {Cui}, Weiguang and {De Petris}, Marco and {Dolag}, Klaus and {Eckert}, Dominique and {Ettori}, Stefano and {Gaspari}, Massimo},
        title = "{The Three Hundred Project: Modeling baryon and hot-gas fraction evolution in simulated clusters}",
      journal = {\aap},
         year = 2025,
        month = oct,
       volume = {702},
          eid = {A182},
        pages = {A182},
          doi = {10.1051/0004-6361/202554283},
archivePrefix = {arXiv},
       eprint = {2505.21624},
 primaryClass = {astro-ph.CO},
       adsurl = {https://ui.adsabs.harvard.edu/abs/2025A&A...702A.182R}
}

@ARTICLE{KnollmanKnebe2009,
       author = {{Knollmann}, S.~R. and {Knebe}, A.},
        title = "{AHF: AMIGA's Halo Finder}",
      journal = {\apjs},
         year = 2009,
        month = may,
       volume = {182},
       number = {2},
        pages = {608-624},
          doi = {10.1088/0067-0049/182/2/608},
       adsurl = {https://ui.adsabs.harvard.edu/abs/2009ApJS..182..608K}
}

@ARTICLE{Walker2019,
       author = {{Walker}, S.~A. and {Simionescu}, A. and {Nagai}, D. and {Okabe}, N. and {Eckert}, D. and {Mroczkowski}, T. and {Akamatsu}, H. and {Ettori}, S. and {Ghirardini}, V.},
        title = "{The Physics of Galaxy Cluster Outskirts}",
      journal = {\ssr},
         year = 2019,
        month = jan,
       volume = {215},
       number = {1},
          eid = {7},
        pages = {7},
          doi = {10.1007/s11214-018-0572-8},
       adsurl = {https://ui.adsabs.harvard.edu/abs/2019SSRv..215....7W}
}

@ARTICLE{Zarattini2025,
       author = {{Zarattini}, S. and {Andreon}, S. and {Puddu}, E.},
        title = "{Position of X-ray low surface brightness clusters in the cosmic filament network}",
      journal = {\aap},
         year = 2025,
        month = feb,
       volume = {694},
          eid = {A256},
        pages = {A256},
          doi = {10.1051/0004-6361/202452744},
       adsurl = {https://ui.adsabs.harvard.edu/abs/2025A&A...694A.256Z}
}

@ARTICLE{Rasia2015,
       author = {{Rasia}, E. and {Borgani}, S. and {Murante}, G. and {Planelles}, S. and {Beck}, A.~M. and {Biffi}, V. and {Ragone-Figueroa}, C. and {Granato}, G.~L. and {Steinborn}, L.~K. and {Dolag}, K.},
        title = "{Cool Core Clusters from Cosmological Simulations}",
      journal = {\apjl},
         year = 2015,
        month = oct,
       volume = {813},
       number = {1},
          eid = {L17},
        pages = {L17},
          doi = {10.1088/2041-8205/813/1/L17},
       adsurl = {https://ui.adsabs.harvard.edu/abs/2015ApJ...813L..17R}
}

@ARTICLE{Springel2003,
       author = {{Springel}, V. and {Hernquist}, L.},
        title = "{Cosmological smoothed particle hydrodynamics simulations: a hybrid multiphase model for star formation}",
      journal = {\mnras},
         year = 2003,
        month = mar,
       volume = {339},
       number = {2},
        pages = {289-311},
          doi = {10.1046/j.1365-8711.2003.06206.x},
       adsurl = {https://ui.adsabs.harvard.edu/abs/2003MNRAS.339..289S}
}

@ARTICLE{Tornatore2007,
       author = {{Tornatore}, L. and {Borgani}, S. and {Dolag}, K. and {Matteucci}, F.},
        title = "{Chemical enrichment of galaxy clusters from hydrodynamical simulations}",
      journal = {\mnras},
         year = 2007,
        month = nov,
       volume = {382},
       number = {3},
        pages = {1050-1072},
          doi = {10.1111/j.1365-2966.2007.12456.x},
       adsurl = {https://ui.adsabs.harvard.edu/abs/2007MNRAS.382.1050T}
}

@ARTICLE{Biffi2017,
       author = {{Biffi}, V. and {Planelles}, S. and {Borgani}, S. and {Fabjan}, D. and {Rasia}, E. and {Murante}, G. and {Tornatore}, L. and {Dolag}, K. and {Granato}, G.~L. and {Gaspari}, M. and {Beck}, A.~M.},
        title = "{The history of chemical enrichment in the intracluster medium from cosmological simulations}",
      journal = {\mnras},
         year = 2017,
        month = jun,
       volume = {468},
       number = {1},
        pages = {531-548},
          doi = {10.1093/mnras/stx444},
       adsurl = {https://ui.adsabs.harvard.edu/abs/2017MNRAS.468..531B}
}

@article{Biffi2018,
   title={The origin of ICM enrichment in the outskirts of present-day galaxy clusters from cosmological hydrodynamical simulations},
   volume={476},
   ISSN={1365-2966},
   url={http://dx.doi.org/10.1093/mnras/sty363},
   DOI={10.1093/mnras/sty363},
   number={2},
   journal={\mnras},
   publisher={Oxford University Press (OUP)},
   author={Biffi, V and Planelles, S and Borgani, S and Rasia, E and Murante, G and Fabjan, D and Gaspari, M},
   year={2018},
   month=feb, pages={2689–2703} }

@article{Kaiser1986,
  author  = {Kaiser, Nick},
  title   = {{Evolution and clustering of rich clusters}},
  journal = {\mnras},
  year    = {1986},
  volume  = {222},
  number  = {2},
  pages   = {323--345},
  month   = {sep},
  doi     = {10.1093/mnras/222.2.323},
  url     = {http://dx.doi.org/10.1093/mnras/222.2.323}
}

@article{Pratt2006,
  author  = {Pratt, G. W. and Arnaud, M. and Pointecouteau, E.},
  title   = {{Gas Entropy in Nearby Galaxy Clusters}},
  journal = {Astronomy \& Astrophysics}, 
  year    = {2006},
  volume  = {446},
  pages   = {429--437},
  doi     = {10.1051/0004-6361:20053525},
  url     = {https://doi.org/10.1051/0004-6361:20053525},
}

@article{Sunyaev1970,
    author = {{Sunyaev}, R.~A. and {Zeldovich}, {Ya}.~B.},
    title = "{The Interaction of Matter and Radiation in a Hot-Model Universe}",
    journal = {Astrophysics and Space Science},
    year = 1970,
    month = apr,
    volume = {7},
    number = {1},
    pages = {3-19},
    doi = {10.1007/BF00653471},
    adsurl = {https://ui.adsabs.harvard.edu/abs/1970Ap&SS...7....3S}
}

@article{Eckert2019,
  author        = {{Eckert}, D. and {Ghirardini}, V. and {Ettori}, S. and {Rasia}, E. and {Biffi}, V. and {Pointecouteau}, E. and {Rossetti}, M. and {Molendi}, S. and {Vazza}, F. and {Gastaldello}, F. and {Gaspari}, M. and {De Grandi}, S. and {Ghizzardi}, S. and {Bourdin}, H. and {Tchernin}, C. and {Roncarelli}, M.},
  title         = "{Non-thermal pressure support in X-COP galaxy clusters}",
  journal       = {\aap},
  year          = 2019,
  month         = jan,
  volume        = {621},
  eid           = {A40},
  pages         = {A40},
  doi           = {10.1051/0004-6361/201833324},
  archivePrefix = {arXiv},
  eprint        = {1810.05165},
  primaryClass  = {astro-ph.HE},
  adsurl        = {https://ui.adsabs.harvard.edu/abs/2019A&A...621A..40E}
}

@article{Lovisari2020,
  author        = {{Lovisari}, L. and {Schellenberger}, G. and {Sereno}, M. and {Ettori}, S. and {Pratt}, G.~W. and {Forman}, W.~R. and {Jones}, C. and {Andrade-Santos}, F. and {Randall}, S. and {Kraft}, R.},
  title         = "{X-Ray Scaling Relations for a Representative Sample of Planck-selected Clusters Observed with XMM-Newton}",
  journal       = {\apj},
  year          = 2020,
  month         = apr,
  volume        = {892},
  number        = {2},
  eid           = {102},
  pages         = {102},
  doi           = {10.3847/1538-4357/ab7997},
  archivePrefix = {arXiv},
  eprint        = {2005.00010},
  primaryClass  = {astro-ph.CO},
  adsurl        = {https://ui.adsabs.harvard.edu/abs/2020ApJ...892..102L}
}

@article{Mantz2021,
  author        = {{Mantz}, A.~B. and {Morris}, R.~G. and {Allen}, S.~W. and {Canning}, R.~E.~A. and {Baumont}, L. and {Benson}, B. and {Bleem}, L.~E. and {Ehlert}, S.~R. and {Floyd}, B. and {Herbonnet}, R. and {Kelly}, P.~L. and {Liang}, S. and {von der Linden}, A. and {McDonald}, M. and {Rapetti}, D.~A. and {Schmidt}, R.~W. and {Werner}, N. and {Wright}, A.},
  title         = "{Cosmological constraints from gas mass fractions of massive, relaxed galaxy clusters}",
  journal       = {\mnras},
  year          = 2021,
  month         = nov,
  volume        = {510},
  number        = {1},
  pages         = {131-145},
  doi           = {10.1093/mnras/stab3390},
  archivePrefix = {arXiv},
  eprint        = {2111.09343},
  primaryClass  = {astro-ph.CO},
  adsurl        = {https://ui.adsabs.harvard.edu/abs/2022MNRAS.510..131M}
}

@article{MDPL2_2016,
  author = {{Klypin}, Anatoly and {Yepes}, Gustavo and {Gottl{\"o}ber}, Stefan and {Prada}, Francisco and {He{\ss}}, Steffen},
        title = "{MultiDark simulations: the story of dark matter halo concentrations and density profiles}",
      journal = {MNRAS},
         year = 2016,
        month = apr,
       volume = {457},
       number = {4},
        pages = {4340-4359},
          doi = {10.1093/mnras/stw248},
archivePrefix = {arXiv},
       eprint = {1411.4001},
 primaryClass = {astro-ph.CO},
       adsurl = {https://ui.adsabs.harvard.edu/abs/2016MNRAS.457.4340K}
}

@article{Pratt_2010,
  title     = {Planck early results. {X}. {Structural} properties of galaxy clusters in the {Planck} {ESZ} sample},
  author    = {Pratt, G. W. and Arnaud, M. and Pointecouteau, E. and Aghanim, N. and Ade, P. A. R. and Aghanim, N. and Armitage-Caplan, C. and Arnaud, M. and Ashdown, M. and others},
  journal   = {Astronomy \& Astrophysics},
  volume    = {536},
  pages     = {A10},
  year      = {2011},
  month     = dec,
  doi       = {10.1051/0004-6361/201116459},
  url       = {https://doi.org/10.1051/0004-6361/201116459},
  publisher = {EDP Sciences}
}

@article{Aguena_2026_ACTDR6,
  title     = {The {Atacama Cosmology Telescope}: {DR6} {Sunyaev-Zel’dovich} Selected Galaxy Clusters Catalog},
  author    = {Aguena, M. and Aiola, S. and Allam, S. and Andrade-Oliveira, F. and Bacon, D. and Bahcall, N. and Battaglia, N. and others},
  journal   = {The Open Journal of Astrophysics},
  volume    = {9},
  year      = {2026},
  month     = jan,
  doi       = {10.33232/001c.155863},
  url       = {https://doi.org/10.33232/001c.155863},
  issn      = {2565-6120},
  publisher = {Maynooth University}
}

@article{LeBrun_2014,
  title     = {Towards a realistic population of simulated galaxy groups and clusters},
  author    = {Le Brun, Amandine M. C. and McCarthy, Ian G. and Schaye, Joop and Ponman, Trevor J.},
  journal   = {\mnras},
  volume    = {441},
  number    = {2},
  pages     = {1270--1290},
  year      = {2014},
  month     = may,
  doi       = {10.1093/mnras/stu608},
  url       = {https://doi.org/10.1093/mnras/stu608},
  publisher = {Oxford University Press}
}

@article{Sembolini_2013,
  title     = {The {MUSIC} of galaxy clusters – {I}. {Baryon} properties and scaling relations of the thermal {Sunyaev–Zel'dovich} effect},
  author    = {Sembolini, F. and Yepes, G. and De Petris, M. and Gottlöber, S. and Hernández-Monteagudo, C. and Puchwein, E.},
  journal   = {\mnras},
  volume    = {429},
  number    = {1},
  pages     = {323--343},
  year      = {2013},
  month     = feb,
  doi       = {10.1093/mnras/sts338},
  url       = {https://doi.org/10.1093/mnras/sts338},
  publisher = {Oxford University Press}
}

@article{Dolag_2016,
  title     = {{SZ} effects in the {Magneticum Pathfinder} simulation: comparison with the {Planck}, {SPT}, and {ACT} results},
  author    = {Dolag, K. and Komatsu, E. and Sunyaev, R.},
  journal   = {\mnras},
  volume    = {463},
  number    = {2},
  pages     = {1797--1811},
  year      = {2016},
  month     = oct,
  doi       = {10.1093/mnras/stw2035},
  url       = {https://doi.org/10.1093/mnras/stw2035},
  publisher = {Oxford University Press}
}

@ARTICLE{Beck2016,
       author = {{Beck}, A.~M. and {Murante}, G. and {Arth}, A. and {Remus}, R.-S. and {Teklu}, A.~F. and {Donnert}, J.~M.~F. and {Planelles}, S. and {Beck}, M.~C. and {Foerster}, P. and {Imgrund}, M. and {Dolag}, K. and {Borgani}, S.},
        title = "{An improved SPH scheme for cosmological simulations}",
      journal = {\mnras},
         year = 2016,
        month = jan,
       volume = {455},
       number = {2},
        pages = {2110-2130},
          doi = {10.1093/mnras/stv2443},
archivePrefix = {arXiv},
       eprint = {1502.07358},
 primaryClass = {astro-ph.CO},
       adsurl = {https://ui.adsabs.harvard.edu/abs/2016MNRAS.455.2110B}
}

@ARTICLE{Cui2022,
       author = {{Cui}, Weiguang and {Dav{\'e}}, Romeel and {Knebe}, Alexander and {Rasia}, Elena and {Gray}, Meghan and {Pearce}, Frazer and {Power}, Chris and {Yepes}, Gustavo and {Anbajagane}, Dhayaa and {Ceverino}, Daniel and {Contreras-Santos}, Ana and {de Andres}, Daniel and {De Petris}, Marco and {Ettori}, Stefano and {Haggar}, Roan and {Li}, Qingyang and {Wang}, Yang and {Yang}, Xiaohu and {Borgani}, Stefano and {Dolag}, Klaus and {Zu}, Ying and {Kuchner}, Ulrike and {Ca{\~n}as}, Rodrigo and {Ferragamo}, Antonio and {Gianfagna}, Giulia},
        title = "{The Three Hundred project: The GIZMO-SIMBA run}",
      journal = {\mnras},
         year = 2022,
        month = jul,
       volume = {514},
       number = {1},
        pages = {977-996},
          doi = {10.1093/mnras/stac1402},
archivePrefix = {arXiv},
       eprint = {2205.03431},
 primaryClass = {astro-ph.CO},
       adsurl = {https://ui.adsabs.harvard.edu/abs/2022MNRAS.514..977C}
}

@ARTICLE{PlanckSZ2015,
       author = {{Planck Collaboration} and {Ade}, P.~A.~R. and {Aghanim}, N. and {Armitage-Caplan}, C. and {Arnaud}, M. and {Ashdown}, M. and {Atrio-Barandela}, F. and {Aumont}, J. and {Aussel}, H. and {Baccigalupi}, C. and {Banday}, A.~J. and {Barreiro}, R.~B. and {Barrena}, R. and {Bartelmann}, M. and {Bartlett}, J.~G. and {Battaner}, E. and {Benabed}, K. and {Beno{\^\i}t}, A. and {Benoit-L{\'e}vy}, A. and {Bernard}, J.-P. and {Bersanelli}, M. and {Bielewicz}, P. and {Bikmaev}, I. and {Bobin}, J. and {Bock}, J.~J. and {B{\"o}hringer}, H. and {Bonaldi}, A. and {Bond}, J.~R. and {Borrill}, J. and {Bouchet}, F.~R. and {Bridges}, M. and {Bucher}, M. and {Burenin}, R. and {Burigana}, C. and {Butler}, R.~C. and {Cardoso}, J.-F. and {Carvalho}, P. and {Catalano}, A. and {Challinor}, A. and {Chamballu}, A. and {Chary}, R.-R. and {Chen}, X. and {Chiang}, H.~C. and {Chiang}, L.-Y. and {Chon}, G. and {Christensen}, P.~R. and {Churazov}, E. and {Church}, S. and {Clements}, D.~L. and {Colombi}, S. and {Colombo}, L.~P.~L. and {Comis}, B. and {Couchot}, F. and {Coulais}, A. and {Crill}, B.~P. and {Curto}, A. and {Cuttaia}, F. and {Da Silva}, A. and {Dahle}, H. and {Danese}, L. and {Davies}, R.~D. and {Davis}, R.~J. and {de Bernardis}, P. and {de Rosa}, A. and {de Zotti}, G. and {Delabrouille}, J. and {Delouis}, J.-M. and {D{\'e}mocl{\`e}s}, J. and {D{\'e}sert}, F.-X. and {Dickinson}, C. and {Diego}, J.~M. and {Dolag}, K. and {Dole}, H. and {Donzelli}, S. and {Dor{\'e}}, O. and {Douspis}, M. and {Dupac}, X. and {Efstathiou}, G. and {En{\ss}lin}, T.~A. and {Eriksen}, H.~K. and {Feroz}, F. and {Ferragamo}, A. and {Finelli}, F. and {Flores-Cacho}, I. and {Forni}, O. and {Frailis}, M. and {Franceschi}, E. and {Fromenteau}, S. and {Galeotta}, S. and {Ganga}, K. and {G{\'e}nova-Santos}, R.~T. and {Giard}, M. and {Giardino}, G. and {Gilfanov}, M. and {Giraud-H{\'e}raud}, Y. and {Gonz{\'a}lez-Nuevo}, J. and {G{\'o}rski}, K.~M. and {Grainge}, K.~J.~B. and {Gratton}, S. and {Gregorio}, A. and {Groeneboom}, E., N. and {Gruppuso}, A. and {Hansen}, F.~K. and {Hanson}, D. and {Harrison}, D. and {Hempel}, A. and {Henrot-Versill{\'e}}, S. and {Hern{\'a}ndez-Monteagudo}, C. and {Herranz}, D. and {Hildebrandt}, S.~R. and {Hivon}, E. and {Hobson}, M. and {Holmes}, W.~A. and {Hornstrup}, A. and {Hovest}, W. and {Huffenberger}, K.~M. and {Hurier}, G. and {Hurley-Walker}, N. and {Jaffe}, A.~H. and {Jaffe}, T.~R. and {Jones}, W.~C. and {Juvela}, M. and {Keih{\"a}nen}, E. and {Keskitalo}, R. and {Khamitov}, I. and {Kisner}, T.~S. and {Kneissl}, R. and {Knoche}, J. and {Knox}, L. and {Kunz}, M. and {Kurki-Suonio}, H. and {Lagache}, G. and {L{\"a}hteenm{\"a}ki}, A. and {Lamarre}, J.-M. and {Lasenby}, A. and {Laureijs}, R.~J. and {Lawrence}, C.~R. and {Leahy}, J.~P. and {Leonardi}, R. and {Le{\'o}n-Tavares}, J. and {Lesgourgues}, J. and {Li}, C. and {Liddle}, A. and {Liguori}, M. and {Lilje}, P.~B. and {Linden-V{\o}rnle}, M. and {L{\'o}pez-Caniego}, M. and {Lubin}, P.~M. and {Mac{\'\i}as-P{\'e}rez}, J.~F. and {MacTavish}, C.~J. and {Maffei}, B. and {Maino}, D. and {Mandolesi}, N. and {Maris}, M. and {Marshall}, D.~J. and {Martin}, P.~G. and {Mart{\'\i}nez-Gonz{\'a}lez}, E. and {Masi}, S. and {Massardi}, M. and {Matarrese}, S. and {Matthai}, F. and {Mazzotta}, P. and {Mei}, S. and {Meinhold}, P.~R. and {Melchiorri}, A. and {Melin}, J.-B. and {Mendes}, L. and {Mennella}, A. and {Migliaccio}, M. and {Mikkelsen}, K. and {Mitra}, S. and {Miville-Desch{\^e}nes}, M.-A. and {Moneti}, A. and {Montier}, L. and {Morgante}, G. and {Mortlock}, D. and {Munshi}, D. and {Murphy}, J.~A. and {Naselsky}, P. and {Nastasi}, A. and {Nati}, F. and {Natoli}, P. and {Nesvadba}, N.~P.~H. and {Netterfield}, C.~B. and {N{\o}rgaard-Nielsen}, H.~U. and {Noviello}, F. and {Novikov}, D. and {Novikov}, I. and {O'Dwyer}, I.~J. and {Olamaie}, M. and {Osborne}, S. and {Oxborrow}, C.~A. and {Paci}, F. and {Pagano}, L. and {Pajot}, F. and {Paoletti}, D. and {Pasian}, F. and {Patanchon}, G. and {Pearson}, T.~J. and {Perdereau}, O.},
        title = "{Planck 2013 results. XXXII. The updated Planck catalogue of Sunyaev-Zeldovich sources}",
      journal = {A\&A},
         year = 2015,
        month = sep,
       volume = {581},
          eid = {A14},
        pages = {A14},
          doi = {10.1051/0004-6361/201525787},
archivePrefix = {arXiv},
       eprint = {1502.00543},
 primaryClass = {astro-ph.CO},
       adsurl = {https://ui.adsabs.harvard.edu/abs/2015A&A...581A..14P}
}

@article{Arnaud2005,
  author  = {Arnaud, M. and Pointecouteau, E. and Pratt, G. W.},
  title   = "{The structural and scaling properties of nearby galaxy clusters: II. The M-T relation}",
  journal = {\aap},
  year    = 2005,
  volume  = {441},
  number  = {3},
  pages   = {893-903},
  doi     = {10.1051/0004-6361:20052856},
  adsurl  = {https://ui.adsabs.harvard.edu/abs/2005A&A...441..893A}
}

@article{Pratt2009,
  author  = {Pratt, G. W. and Croston, J. H. and Arnaud, M. and B{\"o}hringer, H.},
  title   = "{Galaxy cluster X-ray luminosity scaling relations from a representative local sample (REXCESS)}",
  journal = {\aap},
  year    = 2009,
  volume  = {498},
  number  = {2},
  pages   = {361-378},
  doi     = {10.1051/0004-6361/200810994},
  adsurl  = {https://ui.adsabs.harvard.edu/abs/2009A&A...498..361P}
}

@article{Borgani2004,
  author  = {Borgani, S. and Murante, G. and Springel, V. and Diaferio, A. and Dolag, K. and Moscardini, L. and Tormen, G. and Tornatore, L. and Tozzi, P.},
  title   = "{X-ray properties of galaxy clusters and groups from a cosmological hydrodynamical simulation}",
  journal = {\mnras},
  year    = 2004,
  volume  = {348},
  number  = {3},
  pages   = {1078-1096},
  doi     = {10.1111/j.1365-2966.2004.07431.x},
  adsurl  = {https://ui.adsabs.harvard.edu/abs/2004MNRAS.348.1078B}
}

@article{McCarthy2017,
  author  = {McCarthy, I. G. and Schaye, J. and Bird, S. and Le Brun, A. M. C.},
  title   = "{The BAHAMAS project: calibrated hydrodynamical simulations for large-scale structure cosmology}",
  journal = {\mnras},
  year    = 2017,
  volume  = {465},
  number  = {3},
  pages   = {2936-2965},
  doi     = {10.1093/mnras/stw2792},
  adsurl  = {https://ui.adsabs.harvard.edu/abs/2017MNRAS.465.2936M}
}

@article{Puchwein2008,
  author  = {Puchwein, E. and Sijacki, D. and Springel, V.},
  title   = "{Simulations of AGN Feedback in Galaxy Clusters and Groups: Impact on Gas Fractions and the L_X-T Scaling Relation}",
  journal = {\apj},
  year    = 2008,
  volume  = {687},
  number  = {2},
  pages   = {L53-L56},
  doi     = {10.1086/593352},
  adsurl  = {https://ui.adsabs.harvard.edu/abs/2008ApJ...687L..53P}
}

@article{Aihara2022,
  author  = {Aihara, H. and AlSayyad, Y. and Ando, M. and Armstrong, R. and Bosch, J. and Egami, E. and Furusawa, H. and Furusawa, J. and Harasawa, S. and Harikane, Y. and others},
  title   = "{Third data release of the Hyper Suprime-Cam Subaru Strategic Program}",
  journal = {\pasj},
  year    = 2022,
  volume  = {74},
  number  = {2},
  pages   = {247-272},
  doi     = {10.1093/pasj/psab122},
  adsurl  = {https://ui.adsabs.harvard.edu/abs/2022PASJ...74..247A}
}

@ARTICLE{Andreon2014,
       author = {{Andreon}, S.},
        title = "{The important role of evolution in the Planck Y$_{SZ}$-mass calibration}",
      journal = {\aap},
         year = 2014,
        month = oct,
       volume = {570},
          eid = {L10},
        pages = {L10},
          doi = {10.1051/0004-6361/201424387},
archivePrefix = {arXiv},
       eprint = {1409.2987},
 primaryClass = {astro-ph.CO}
}

@ARTICLE{Planck2014_XX,
       author = {{Planck Collaboration} and {Ade}, P.~A.~R. and {Aghanim}, N. and {Armitage-Caplan}, C. and {Arnaud}, M. and {Ashdown}, M. and {Atrio-Barandela}, F. and {Aumont}, J. and {Baccigalupi}, C. and {Banday}, A.~J. and et al.},
        title = "{Planck 2013 results. XX. Cosmology from Sunyaev-Zeldovich cluster counts}",
      journal = {\aap},
         year = 2014,
        month = nov,
       volume = {571},
          eid = {A20},
        pages = {A20},
          doi = {10.1051/0004-6361/201321521},
archivePrefix = {arXiv},
       eprint = {1303.5080},
 primaryClass = {astro-ph.CO}
}

@article{Gianfagna_2022,
   title   = {A study of the hydrostatic mass bias dependence and evolution within {The Three Hundred} clusters},
   author  = {Gianfagna, Giulia and Rasia, Elena and Cui, Weiguang and De Petris, Marco and Yepes, Gustavo and Contreras-Santos, Ana and Knebe, Alexander},
   journal = {\mnras},
   year    = {2022},
   volume  = {518},
   number  = {3},
   pages   = {4238--4248},
   doi     = {10.1093/mnras/stac3364},
   url     = {https://doi.org/10.1093/mnras/stac3364},
   month   = {nov}
}

@article{Press1974,
  author    = {{Press}, William H. and {Schechter}, Paul},
  title     = "{Formation of Galaxies and Clusters of Galaxies by Self-Similar Gravitational Condensation}",
  journal   = {The Astrophysical Journal},
  year      = {1974},
  month     = {feb},
  volume    = {187},
  pages     = {425-438},
  doi       = {10.1086/152650},
  url       = {http://dx.doi.org/10.1086/152650},
  issn      = {1538-4357},
  publisher = {American Astronomical Society}
}

@ARTICLE{Sousbie2011,
       author = {{Sousbie}, T.},
        title = "{The persistent cosmic web and its filamentary structure - I. Theory and implementation}",
      journal = {\mnras},
         year = 2011,
        month = jun,
       volume = {414},
        number = {1},
        pages = {350-383},
          doi = {10.1111/j.1365-2966.2011.18394.x},
          url = {https://doi.org/10.1111/j.1365-2966.2011.18394.x}
}

@ARTICLE{Bond1996,
       author = {{Bond}, J. Richard and {Kofman}, Lev and {Pogosyan}, Dmitry},
        title = "{How filaments of galaxies are woven into the cosmic web}",
      journal = {\nat},
         year = 1996,
        month = apr,
       volume = {380},
        number = {6575},
        pages = {603-606},
          doi = {10.1038/380603a0},
          url = {https://doi.org/10.1038/380603a0}
}

@ARTICLE{Codis2018,
       author = {{Codis}, Sandrine and {Pogosyan}, Dmitri and {Pichon}, Christophe},
        title = "{On the connectivity of the cosmic web: theory and implications for cosmology and galaxy formation}",
      journal = {\mnras},
         year = 2018,
        month = jun,
       volume = {479},
        number = {1},
        pages = {973-993},
          doi = {10.1093/mnras/sty1643},
          url = {https://doi.org/10.1093/mnras/sty1643}
}

@article{McCarthy2011,
  author   = {McCarthy, I. G. and Schaye, J. and Bower, R. G. and Ponman, T. J. and Booth, C. M. and Dalla Vecchia, C. and Springel, V.},
  title    = {Gas expulsion by quasar-driven winds as a solution to the overcooling problem in galaxy groups and clusters: how quasar winds solve overcooling in groups},
  journal  = {\mnras},
  year     = {2011},
  volume   = {412},
  number   = {3},
  pages    = {1965--1984},
  month    = jan,
  doi      = {10.1111/j.1365-2966.2010.18033.x},
  url      = {https://doi.org/10.1111/j.1365-2966.2010.18033.x}
}

@article{Barnes2017,
  author   = {Barnes, David J. and Kay, Scott T. and Bah{\'e}, Yannick M. and Dalla Vecchia, Claudio and McCarthy, Ian G. and Schaye, Joop and Bower, Richard G. and Jenkins, Adrian and Thomas, Peter A. and Schaller, Matthieu and Crain, Robert A. and Theuns, Tom and White, Simon D. M.},
  title    = {The Cluster-EAGLE project: global properties of simulated clusters with resolved galaxies},
  journal  = {\mnras},
  year     = {2017},
  volume   = {471},
  number   = {1},
  pages    = {1088--1106},
  month    = jun,
  doi      = {10.1093/mnras/stx1647},
  url      = {https://doi.org/10.1093/mnras/stx1647}
}

@article{Zhang2026,
  author        = {Zhang, Zhao Joseph and Nagamine, Kentaro and Oku, Yuri and Lee, Khee-Gan and Fukushima, Keita and Tomaru, Kazuki and Zhang, Bing and Medlock, Isabel and Nagai, Daisuke},
  title         = {Probing the Cosmic Baryon Distribution and the Impact of Active Galactic Nuclei Feedback with Fast Radio Bursts in {CROCODILE} Simulation},
  journal       = {arXiv e-prints},
  year          = {2026},
  month         = mar,
  eprint        = {2503.12741},
  archivePrefix = {arXiv},
  primaryClass  = {astro-ph.CO},
  doi           = {10.48550/arXiv.2503.12741},
  url           = {https://arxiv.org/abs/2503.12741}
}

@article{Santoni2024,
  author   = {Santoni, Sara and De Petris, Marco and Yepes, Gustavo and Ferragamo, Antonio and Bianconi, Matteo and Gray, Meghan E. and Kuchner, Ulrike and Pearce, Frazer R. and Cui, Weiguang and Ettori, Stefano},
  title    = {{The THREE HUNDRED project: Estimating the dependence of gas filaments on the mass of galaxy clusters}},
  journal  = {Astronomy \& Astrophysics},
  year     = {2024},
  volume   = {692},
  pages    = {A44},
  month    = dec,
  doi      = {10.1051/0004-6361/202450895},
  url      = {https://doi.org/10.1051/0004-6361/202450895}
}

@article{Oguri2021,
  title   = {Hundreds of weak lensing shear-selected clusters from the Hyper Suprime-Cam Subaru Strategic Program S19A data},
  author  = {Oguri, Masamune and Miyazaki, Satoshi and Li, Xiangchong and Luo, Wentao and Mitsuishi, Ikuyuki and Miyatake, Hironao and More, Surhud and Nishizawa, Atsushi J and Okabe, Nobuhiro and Ota, Naomi and {Plazas Malag{\'o}n}, Andr{\'e}s A and Utsumi, Yousuke},
  journal = {Publications of the Astronomical Society of Japan},
  volume  = {73},
  number  = {4},
  pages   = {817-829},
  year    = {2021},
  month   = {may},
  doi     = {10.1093/pasj/psab047},
  url     = {http://dx.doi.org/10.1093/pasj/psab047},
  issn    = {2053-051X}
}

@article{Ragagnin_2022,
  title     = {Simulation view of galaxy clusters with low X-ray surface brightness},
  author    = {Ragagnin, A. and Andreon, S. and Puddu, E.},
  journal   = {Astronomy \& Astrophysics},
  volume    = {666},
  pages     = {A22},
  year      = {2022},
  month     = {sep},
  publisher = {EDP Sciences},
  doi       = {10.1051/0004-6361/202244397},
  url       = {http://dx.doi.org/10.1051/0004-6361/202244397},
  issn      = {1432-0746}
}

@article{Vikhlinin_2009,
  title     = {{\textit{CHANDRA}} CLUSTER COSMOLOGY PROJECT. II. SAMPLES AND X-RAY DATA REDUCTION},
  author    = {Vikhlinin, A. and Burenin, R. A. and Ebeling, H. and Forman, W. R. and Hornstrup, A. and Jones, C. and Kravtsov, A. V. and Murray, S. S. and Nagai, D. and Quintana, H. and Voevodkin, A.},
  journal   = {The Astrophysical Journal},
  volume    = {692},
  number    = {2},
  pages     = {1033--1059},
  year      = {2009},
  month     = {feb},
  publisher = {American Astronomical Society},
  doi       = {10.1088/0004-637X/692/2/1033},
  url       = {http://dx.doi.org/10.1088/0004-637X/692/2/1033},
  issn      = {1538-4357}
}

@article{Pacaud2007,
  title     = {The {\textit{XMM}}-LSS survey: the Class 1 cluster sample over the initial 5 deg$^{2}$ and its cosmological modelling},
  author    = {Pacaud, F. and Pierre, M. and Adami, C. and Altieri, B. and Andreon, S. and Chiappetti, L. and Detal, A. and Duc, P.-A. and Galaz, G. and Gueguen, A. and Le F{\`e}vre, J.-P. and Hertling, G. and Libbrecht, C. and Melin, J.-B. and Ponman, T. J. and Quintana, H. and Refregier, A. and Sprimont, P.-G. and Surdej, J. and Valtchanov, I. and Willis, J. P. and Alloin, D. and Birkinshaw, M. and Bremer, M. N. and Garcet, O. and Jean, C. and Jones, L. R. and Le F{\`e}vre, O. and Maccagni, D. and Mazure, A. and Proust, D. and R{\"o}ttgering, H. J. A. and Trinchieri, G.},
  journal   = {Monthly Notices of the Royal Astronomical Society},
  volume    = {382},
  number    = {3},
  pages     = {1289--1308},
  year      = {2007},
  month     = {nov},
  publisher = {Oxford University Press (OUP)},
  doi       = {10.1111/j.1365-2966.2007.12468.x},
  url       = {http://dx.doi.org/10.1111/j.1365-2966.2007.12468.x},
  issn      = {1365-2966}
}

% %%%%%%%%%%%%%%%%%%%%%%%%%%%%%%%%%%%%%%%%%%%%%%%%%%%%%%%%%%%%%%
% % WARNING
% % Please note that we have included the references below in
% % order to compile the document, but we ask you to:
% %
% % - use BibTeX with the regular commands:
% %   \bibliographystyle{aa} % style aa.bst
% %   \bibliography{Yourfile} % your references Yourfile.bib
% % - join the .bib files when you upload your source files
% %%%%%%%%%%%%%%%%%%%%%%%%%%%%%%%%%%%%%%%%%%%%%%%%%%%%%%%%%%%%%%

% Appendices must be placed after   \end{thebibliography}
% They will be placed automatically on a new page.
%%%%%%%%%%%%%%%%%%%%%%%%%%%%%%%%%%%%%%%%%%%%%%%%%%%%%%%%%%%%%%%
\begin{appendix}

% \section{Model selection and LGFCs selection}
% Table \ref{Valori chi quadro} reports the $\chi^2_{red}$ values for the linear, parabolic, and logarithmic model, at each redshift analyzed. These results clearly indicate that the parabolic model provides the best overall fit to the $f_{g,500} - M_{500}$ relation across all redshifts.
% \begin{table}[ht!]
%     \centering
%     \caption{$\chi^2_{red}$ values for the different best-fitting models at each redshift}
%     \begin{tabular}{c c c c}
%     \hline\hline
%          Redshift & Linear & Parabolic & Logarithmic \\
%          \hline
%          0.817 & 0.48 & 0.01 & 0.19\\
%          0.700 & 0.84 & 0.02 & 0.39\\
%          0.592 & 0.87 & 0.01 & 0.44\\
%          0.490 & 0.51 & 0.08 & 0.20\\
%          0.394 & 0.62 & 0.04 & 0.25\\
%          0.304 & 0.97 & 0.02 & 0.55\\
%          0.116 & 0.68 & 0.05 & 0.33\\
%          0 & 0.59 & 0.05 & 0.27\\
%          \hline
%     \end{tabular}
    
%     \label{Valori chi quadro}
% \end{table}

\section{Redshift evolution of the $f_{\rm gas,500} - M_{500}$ relation}
\begin{figure}[ht!]
    \centering
    \includegraphics[width=8.8cm]{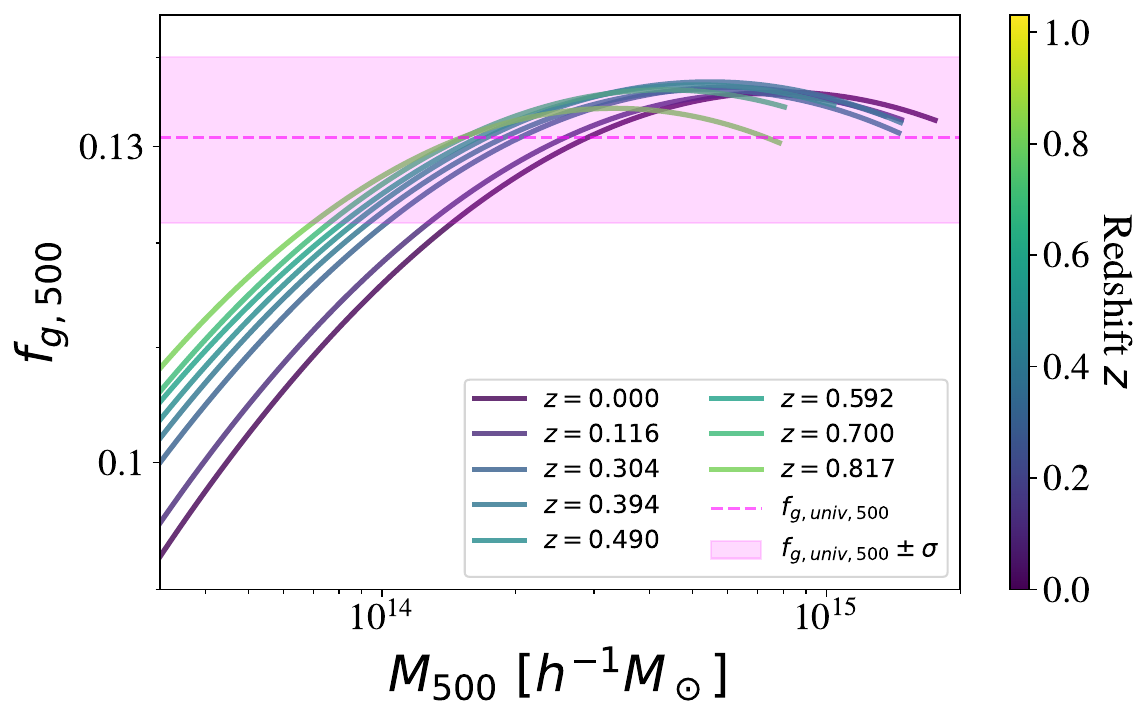}
    \caption{Redshift evolution of $f_{\rm g,500} - M_{500}$. The different curves are the best fit using the parabola model (Eq. \ref{Parabola model}). They are colored depending on the redshift value. The magenta horizontal dashed line represents the universal gas fraction value $f_{g,\mathrm{univ},500}$.}
    \label{fgas profiles redshift evol}
\end{figure}

\begin{figure}[ht!]
    \centering
    \includegraphics[width=8.8cm]{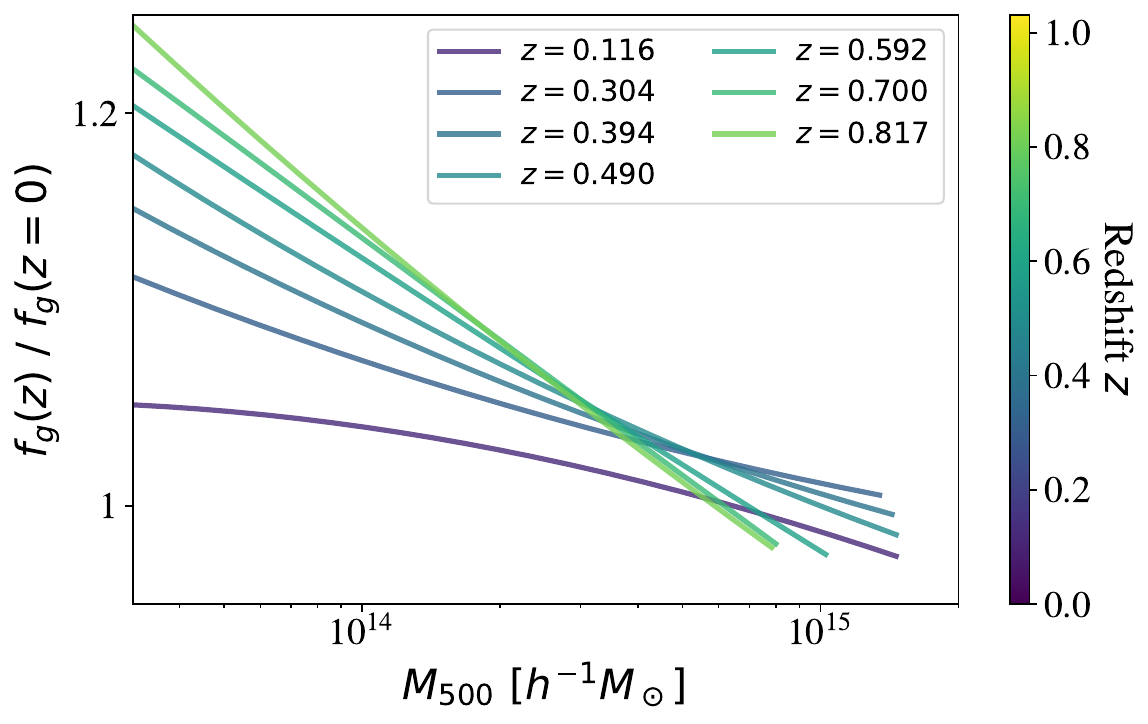}
    \caption{Evolution of $f_{\rm g,500} - M_{500}$ scaling relation at different redshifts, taking $z=0$ as the reference.}
    \label{fg/fg0}
\end{figure}

In Fig. \ref{fgas profiles redshift evol} we report the redshift evolution of the $f_{\rm g,500} - M_{500}$ relation. We report the best-fitting curves at different redshifts, with a range of colors ranging from the darker tones (higher redshifts), to lighter ones (lower redshifts). The horizontal magenta line marks the universal median gas fraction value $f_{\rm g,\mathrm{univ},500} = 0.131$, while the magenta shaded area represents the uncertainty $\sigma = 0.009$, as measured by \cite{Eckert2019}. At fixed mass, the gas fraction decreases with decreasing redshift. The mass range probed by the relation extends toward higher masses as the redshift decreases, reflecting the cosmological hierarchical scenario.
In Fig. \ref{fg/fg0} we show the relative evolution of $f_{\rm g,500}-M_{500}$ at various redshifts compared to that at redshift $z=0$. We find a clear mass-dependent evolution: low mass systems show significantly higher gas fraction values at high redshift compared to their present-day counterparts, with differences reaching up to $\sim +25 \%$ at redshift $z = 0.817$. Differently, massive clusters show little or no evolution. This behavior indicates a progressive depletion in low-mass clusters toward low redshift, likely driven by non-gravitational processes such as AGN feedback and environmental processes. Instead, in high-mass clusters the deeper gravitational potential wells dominate over these processes and we have a self-similar evolution ($f_g \approx \mathrm{const}$).
This trend reflects the scenario described by \cite{Rasia2025}: GCs that are already massive at high redshift ($M_{200} > 2\times 10^{14}\mathrm{M_\odot}$ at $z\geq 1$)  experience an horizontal shift in the $f_{\rm g,500}-M_{500}$ plane conversely to lower-mass systems which undergo mass and structural growth, moving diagonally in the $f_{\rm g,500} - M_{500}$ plane.

\vspace{5cm}
\section{Radial profiles}
\begin{figure*}[tp]
    \centering
    \includegraphics[width=180mm]{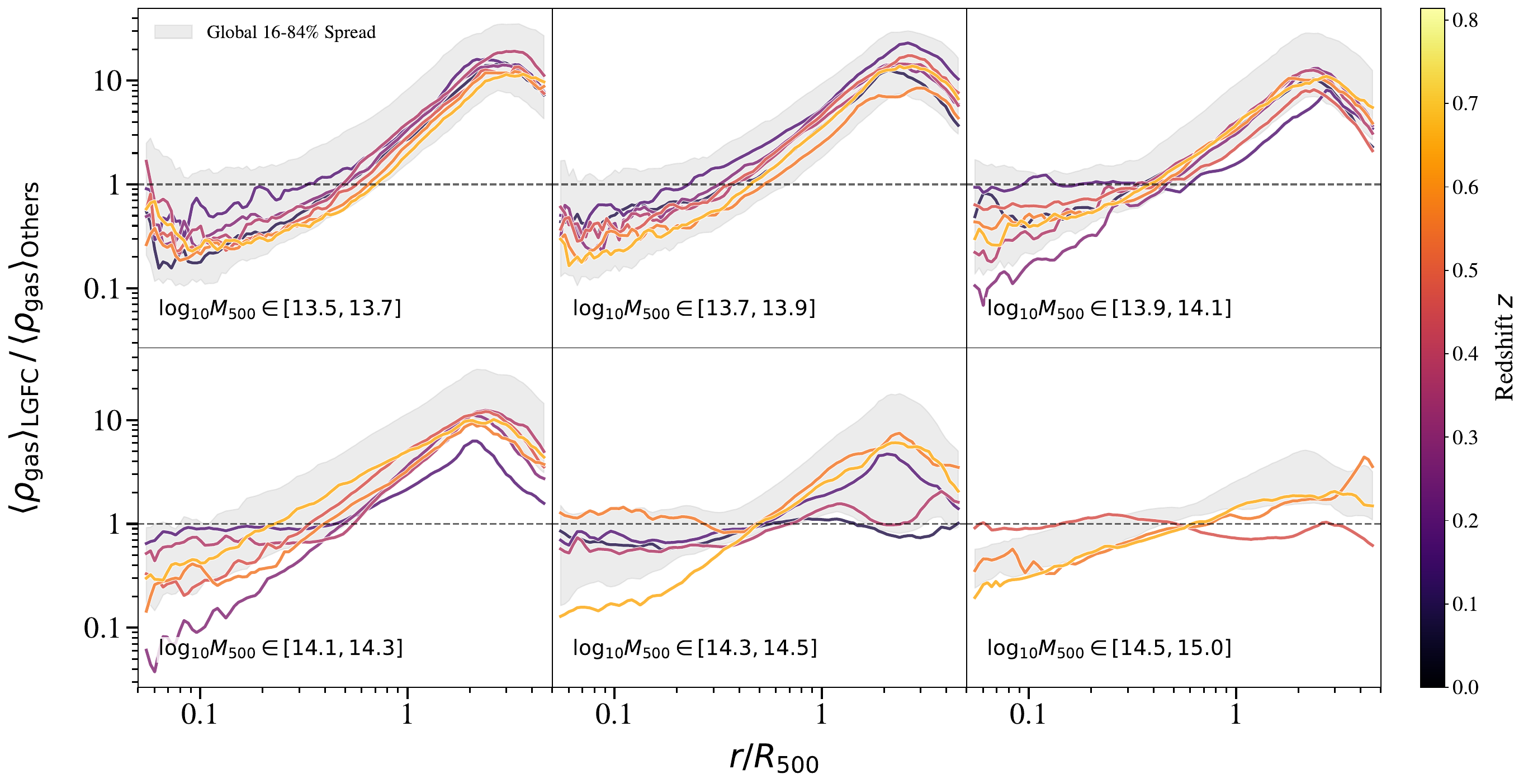}
    \caption{Ratio between the gas density profiles for the LGFCs population and those of the No-LGFCs population. They are colored depending on the redshift value. Gray shaded region represents the intrinsic scatter.}
    \label{Gas density ratio}
\end{figure*}

\begin{figure*}[tp]
    \centering
    \includegraphics[width=180mm]{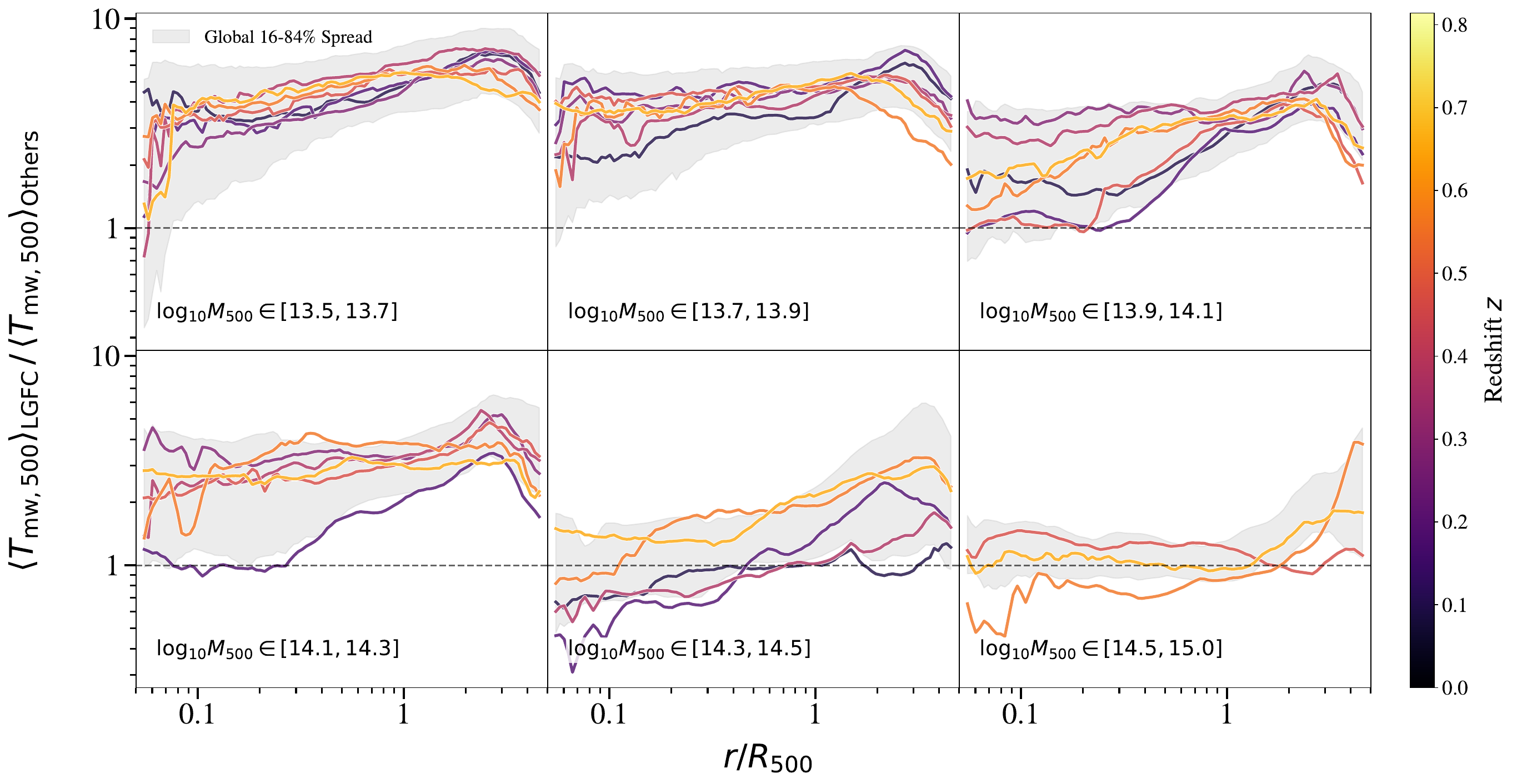}
    \caption{Ratio between the gas temperature profiles for the LGFCs population and those of the No-LGFCs population. They are colored depending on the redshift value. Gray shaded region represents the intrinsic scatter.}
    \label{Gas temperature ratio}
\end{figure*}

We report the ratio of density and temperature median profiles between the two populations in Fig. \ref{Gas density ratio} and in Fig. \ref{Gas temperature ratio}, respectively.

\section{Bias dependence on redshift and total mass}
\begin{figure*}[tp]
    \centering
    \includegraphics[width=150mm]{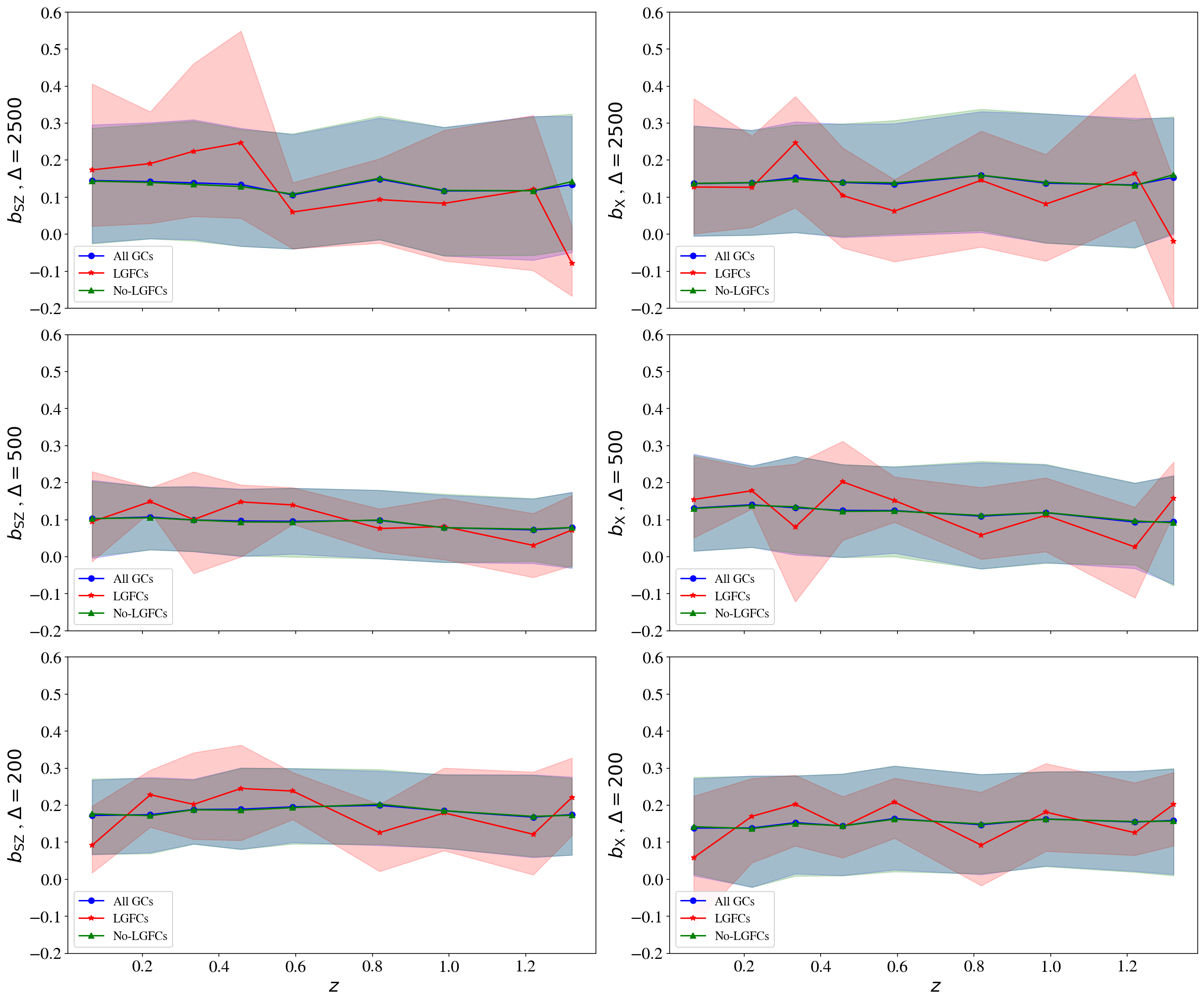}
    \caption{Redshift evolution of the biases $b_{\rm SZ}$ (left-hand panels) and $b_{\rm X}$ (right-hand panels). The bias estimated at $\Delta = 2500, 500$ and $200$ are represented in the top, middle and bottom panels, respectively. The median values of the bias for all GCs, LGFCs and No-LGFCs are represented with blue circles, red stars and green triangles, respectively. The shaded regions represent the 16th-84th percentile values. }
    \label{bias redshift}
\end{figure*}

\begin{figure*}[tp]
    \centering
    \includegraphics[width=150mm]{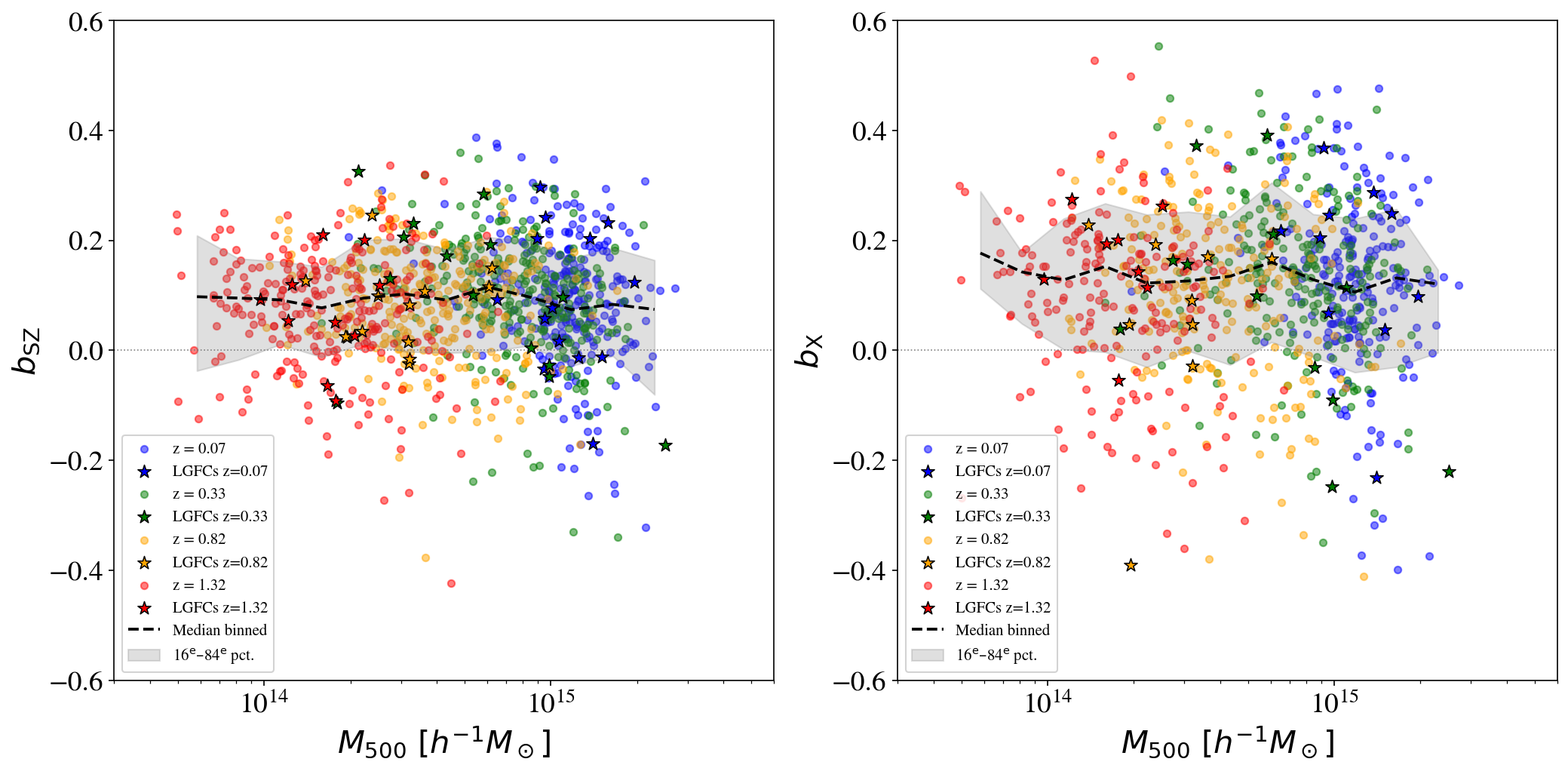}
    \caption{Dependence of $b_{\rm SZ}$ (left panel) and $b_{\rm X}$ (right panel) on the total mass $M_{500}$. The points are colored according to their redshift value: $z = 1.32, 0.82, 0.33, 0.07$ are colored in red, yellow, green and blue, respectively. The black dashed line and the shaded gray area represent the median and the 16th-84th percentiles of the biases in mass bins}
    \label{bias mass}
\end{figure*}

We report the plots regarding the dependence of the $b_{\rm SZ}$ and $b_{\rm X}$ biases on the redshift and on the total mass $M_{500}$ in Fig. \ref{bias redshift} and in Fig. \ref{bias mass}, respectively.


\end{appendix}
\end{document}